\documentclass[prx,aps,superscriptaddress,twocolumn,longbibliography]{revtex4-2}

\usepackage{mathtools}
\usepackage{mathrsfs, amsmath, wasysym}
\usepackage{bbold}
\usepackage[mathscr]{eucal}
\usepackage{xfrac}
\usepackage{graphicx}
\usepackage{dcolumn}
\usepackage{bm}
\usepackage{color}
\usepackage{tabularx}
\usepackage{natbib}
\usepackage[colorlinks,linkcolor=blue,citecolor=blue,urlcolor=blue]{hyperref}
\usepackage[normalem]{ulem}
\usepackage[all]{hypcap}
\usepackage[dvipsnames,usenames,table]{xcolor}
\usepackage{braket,mleftright}

\newcommand{\nocontentsline}[3]{}
\newcommand{\tocless}[2]{\bgroup\let\addcontentsline=\nocontentsline#1{#2}\egroup}

\newcommand{\bk}{{\bf k}}

\newcommand{\bq}{{\bf q}}
\newcommand{\bx}{{\bf x}}
\newcommand{\by}{{\bf y}}
\newcommand{\bz}{{\bf z}}

\newcommand{\ba}{{\bf a}}

\newcommand{\bb}{{\bf b}}
\newcommand{\bB}{{\bf B}}

\newcommand{\bE}{{\bf E}}
\newcommand{\bd}{{\bf d}}

\newcommand{\bP}{{\bf P}}

\newcommand{\bn}{{\bf n}}

\newcommand{\bM}{{\bf M}}

\newcommand{\bh}{{\bf h}}
\newcommand{\bT}{{\bf T}}
\newcommand{\bN}{{\bf N}}

\newcommand{\btau}{\boldsymbol{\tau} }
\newcommand{\bsigma}{\boldsymbol{\sigma} }
\newcommand{\brho}{\boldsymbol{\rho} }

\newcommand{\ehat}{\hat {\bf e}}

\newcommand{\be}{\begin{equation}}
\newcommand{\ee}{\end{equation}}
\newcommand{\beg}{\begin{gather}}
\newcommand{\eeg}{\end{gather}}
\newcommand{\beq}{\begin{eqnarray}}
\newcommand{\eeq}{\end{eqnarray}}
\newcommand{\bea}{\begin{align}}
\newcommand{\eea}{\end{align}}
\newcommand{\beqq}{\begin{eqnarray*}}
\newcommand{\eeqq}{\end{eqnarray*}}

\newcommand{\up}{\uparrow}
\newcommand{\down}{\downarrow}

\begin{document}

\title{Topological spin multipolization and linear magnetoelectric coupling \\
in two-dimensional antiferromagnets}

\author{J\"orn W.~F. Venderbos}
\affiliation{Department of Physics, Drexel University, Philadelphia, PA 19104, USA}
\affiliation{Department of Materials Science \& Engineering, Drexel University, Philadelphia, Pennsylvania 19104, USA \looseness=-1}

\author{Paola Gentile}
\affiliation{CNR-SPIN, I-84084 Fisciano (Salerno), Italy, c/o Universit\'a di Salerno, I-84084 Fisciano (Salerno), Italy}

\author{Carmine Ortix}
\affiliation{Dipartimento di Fisica ``E. R. Caianiello", Universit\`a di Salerno, IT-84084 Fisciano (SA), Italy}
\affiliation{CNR-SPIN, I-84084 Fisciano (Salerno), Italy, c/o Universit\'a di Salerno, I-84084 Fisciano (Salerno), Italy}

\date \today


\begin{abstract}
In this paper we predict that the magnetoelectric response of two-dimensional (2D) antiferromagnets contains a contribution determined by the topological properties of the ground state. 
The quasitopological magnetoelectric response, encoded in the spin magnetoelectric polarizability and the closely related spin multipolization, occurs when the electronic structure of the antiferromagnetic insulator is described by massive 2D Dirac fermions, and is therefore native to 2D, unlike the topological magnetoelectric effect of three-dimensional topological insulators. To demonstrate the topology-induced contribution to the (spin) magnetoelectric polarizability, we compute the magnetoelectric polarizability microscopically for two distinct minimal lattice models: a spin-orbit coupled N\'eel antiferromagnet and a spin-orbit-free noncollinear antiferromagnet with double-$Q$ spin order. We show that the topological origin of the revealed magnetoelectric effect can be traced back to the electromagnetic response of topological semimetals in two dimensions, and hence is ultimately governed by a strong topological invariant in one dimension. Given this dimensional hierarchy, we further consider two minimal lattice models in one dimension, both one-dimensional variants of the 2D lattice models, and show that the magnetoelectric polarizability exhibits a clear signature of nontrivial crystalline topology. Possible material realizations are discussed.
\end{abstract}

\maketitle

\section{Introduction \label{sec:intro}}

The study of magnetoelectric effects in solids constitutes a premier research area within condensed matter physics. Since magnetoelectric effects generally reflect an intricate coupling of distinct ordering phenomena, their observation and understanding provide key insight into the physics of strongly correlated materials~\cite{Fiebig:2005pR123,Spaldin:2005p391}. They are not only of fundamental interest, however, but also of great practical relevance~\cite{Manipatruni2019}. The ability to control magnetic properties of materials with electric fields, or, conversely, electric properties with magnetic fields, is destined to play a key role in future applications, for instance in the field of spintronics~\cite{Zutic:2004p323,Baltz:2018p015005}, where efficient control of magnetism by voltage switches is highly desirable. 

Materials with strong magnetoelectric coupling are colloquially known as magnetoelectrics. A prominent example is the family of multiferroics~\cite{Fiebig:2005pR123,Khomksii:2006p1,Ramesh:2007p21,Cheong:2007p13,Eerenstein:2006p759,Tokura:2014p076501,Barone:2015p143,Fiebig2016,Spaldin2019}. In the more narrow and traditional sense multiferroics are understood as materials in which a ferroelectric polarization spontaneously forms in the magnetically ordered state~\cite{Chupis:1969p2818,Katsura:2004p057205}. This kind of coupling between magnetism and ferroelectric order is enabled by a particular combination of crystal structure (i.e., the symmetry of the crystal) and magnetic structure (i.e., the type of magnetic order): these must satisfy the requirement that a ferroelectric polarization is allowed once magnetic order develops. From a symmetry perspective, the traditional multiferroics are thus magnetically ordered states with an induced electric dipole moment. Clearly, a necessary condition is that magnetic order spontaneously breaks the inversion symmetry of the crystal lattice, and this is known to occur, for instance, in the case of certain helimagnets~\cite{Jablonskii:1983p673,Kimura:2003p55,Hur:2004p392,Hur:2004p107207,Goto:2004p257201,Mostovoy2006,Cheong:2007p13}, or due to the formation of dislocated spin density waves~\cite{Betouras2007}. 

More broadly, the class of magnetoelectrics comprises magnetically ordered states characterized by higher multipole order, such as magnetic quadrupole or magnetic toroidal dipole order, which are each associated with distinct patterns of magnetoelectric response. These higher multipole orders break time-reversal and spatial inversion symmetries, but preserve their product. This forbids ferroelectricity but allows for a linear magnetization ($\bM$) response to an applied electric field ($\bE$) and a linear polarization ($\bP$) response to an applied magnetic field ($\bB$). For instance, in the case of magnetic order characterized by a toroidal moment $\bT$~\cite{Gorbatsevich:1994p321,Kopaev:2009p1111}, the magnetization response is of the form $\bM = \bT \times \bE$ and the polarization response is given by $\bP = \bB \times \bT $. Multiferroics---and more generally magnetoelectrics---hence fit into a broader classification scheme naturally and elegantly framed in terms of the (lowest order) magnetoelectric multipole moment induced in the ordered magnetic state.

Materials exhibiting such multipolar order and the associated magnetoelectric responses have seen a surge in interest~\cite{Spaldin2008}, and have come under scrutiny from a number of different perspectives. From a conceptual perspective, much work has focused on the proper categorization of distinct types of magnets and their magnetoelectric responses~\cite{Hayami:2018p165110,Suzuki:2017p094406,Hayami:2024p072001}. The microscopic perspective has emphasized the importance of properly defining multipole moment densities (e.g., the toroidization or monopolization) as bulk quantities~\cite{Ederer:2007p214404,Spaldin:2013p094429,Thoele:2016p195167,Gao:2018p134423}, leading to a microscopic theory for calculating these quantities in terms of quantum mechanical wave functions. The experimental and materials perspective has focused on finding and characterizing material realizations of higher multipole magnetoelectric coupling~\cite{Aken2007,Lehmann2019,Zimmermann2014,Bhowal:2022p227204,Ortix2023,Pylypovskyi2025}. 

Magnetoelectric responses in solids have also attracted a great deal of attention in a context quite different from unconventional magnetism: the realm of topological insulators. Following their discovery, it was shown that $\mathbb{Z}_2$ topological insulators in three dimensions (3D) are associated with a quantized magnetoelectric polarization $P_3=n+1/2$, where $n$ is an integer, which manifests as a magnetoelectric term in the electromagnetic Lagrangian given by $\mathcal L_{em} = (P_3 /2\pi)\bE\cdot \bB $~\cite{Qi:2008p195424,Essin:2009p146805,Mong:2010p245209,Hughes:2011p245132,Turner:2012p165120,Zhang:2019p206401}. This topological magnetoelectric coupling leads to a number of unusual quantized response phenomena with promising potential for device implementation~\cite{Qi:2008p195424,Maciejko:2010p166803,Rosenberg:2010p035105,Li:2010p284,Tse:2010p057401,Nomura:2011p166802,Morimoto:2015p085113,Wang:2015p081107}. Conceptually, it establishes a deeper connection between magnetoelectric behavior and the topology of the electronic ground state.

In this paper, we consider magnetoelectric coupling in two-dimensional antiferromagnets and demonstrate the existence of a quasitopological contribution to the magnetoelectric effect distinct from the topological magnetoelectric effect in three dimensions. The magnetoelectric effect studied here is native to two-dimensional magnets~\cite{Burch:2018p47,Gibertini:2019p408} and occurs in antiferromagnetic insulators which break space inversion symmetry, but preserve the product of time-reversal and inversion. These systems are classified by a magnetoelectric multipole moment and admit a linear magnetoelectric response. We show that the associated magnetoelectric polarizability, and consequently the multipolization (i.e. the multipole moment densities), can have a contribution of topological origin. This is similar to---but, as mentioned, distinct in nature from---the Chern-Simons term which appears in the (orbital) magnetoelectric polarizability in three dimensions~\cite{Essin:2010p205104,Malashevich:2010p053032}. The topology-induced magnetoelectric coupling discussed here can be traced back to the topology of Dirac fermions, and arises when the antiferromagnetic insulator can be described by a gapped Dirac fermion theory. 

We establish our results by computing the magnetoelectric properties of 2D antiferromagnets microscopically, using formulae for the spin magnetoelectric polarizability and  spin multipolization applicable to periodic crystals. In general, there are two contributions to the magnetoelectric properties of materials: one of spin and one or orbital origin. Here we focus on the spin contribution. Prior work~\cite{Gao:2018p134423} has shown how to define and calculate the spin multipolization, as well as the closely related spin magnetoelectric polarizability, in periodic magnetic systems, and in this work we demonstrate that, when applied to a special class of antiferromagnets, these microscopic expressions carry signatures of nontrivial topology. 

Two complementary minimal lattice models are examined to showcase this general result. The first is a buckled square lattice model of a spin-orbit coupled N\'eel antiferromagnet. Apart from satisfying the symmetry requirements for magnetoelectric coupling, this model derives its broader significance from the crystal symmetry-enforced Dirac points realized in the nonmagnetic state~\cite{Young:2015p126803}. We will show that these are the source of the topological magnetoelectric effect. The second model is a square lattice model of a double-$Q$ non-collinear spin state and additional hopping modulations (i.e., bond density wave). In this model spin-rotation symmetry is broken by the ordered spin configuration, and magnetoelectric coupling occurs even without spin-orbit coupling. The topological magnetoelectric effect arises from the Dirac crossings present in the reconstructed energy bands of the double-$Q$ spin structure~\cite{Agterberg:2000p13816}. 

For both models we develop a Landau theory approach to determining the symmetry-allowed magnetoelectric responses~\cite{Harris:2007p054447,Harris:2012p100403}. This approach takes the high-symmetry nonmagnetic state as a starting point and considers all allowed trilinear or quadrilinear couplings between the magnetic order parameter, any other relevant order parameter (e.g. bond density wave), and the polarization and magnetization. For a given realization of magnetic order, the magnetoelectric tensor can then be directly inferred from these couplings. Since such a Landau theory analysis applies irrespective of the underlying symmetry group, in particular insofar as the inclusion of spin-orbit coupling is concerned, it provides an appealing universally applicable framework for understanding the magnetoelectric phenomena in magnetically ordered systems.

The topology-induced contribution to the (spin) magnetoelectric polarizability and monopolization raises the important question as to what its precise deeper origin is. Is it signaling a strong topological invariant, similar to the magnetoelectric polarization in three dimensions? We address and answer this question by drawing a connection with the electromagnetic response theory of two-dimensional Dirac semimetals~\cite{Ramamurthy:2015p085105}. Since two-dimensional semimetals can be viewed as a collection of one-dimensional insulating phases that undergo a band inversion transition as a function of momentum~\cite{Ryu2002}, their electromagnetic response is governed by a topological invariant in one dimension~\cite{Ramamurthy:2015p085105}. As a result, the topological origin of the magnetoelectric polarizability is ultimately a manifestation of topology in one dimension, more specifically of the associated quantized electronic contribution to the polarization~\cite{Zak1989,Hughes:2011p245132,Lau2015,Miert2017,Miert2017b}. 
It therefore  constitutes a so-called ``quasitopological'' response (see Secs.~\ref{ssec:em-dirac} and \ref{ssec:patterns-EM}). 

To further corroborate and highlight the significance of the topological character in one dimension (1D), we also examine magnetoelectric coupling in 1D toy lattice models of antiferromagnets~\cite{Yanase:2014p014703,Hayami:2015p064717,Hayami:2022p123701,Yatsushiro:2022p155157,Hayami:2023p094106}. Two such models are considered specifically, which each represent 1D variants of the aforementioned two-dimensional models. In parts of the phase diagram where crystalline symmetries such as spatial inversion are preserved, insulating ground states admit a crystalline symmetry-protected $\mathbb{Z}_2$ classification with quantized fractional polarization~\cite{Zak1989,Hughes:2011p245132,Lau2015,Miert2017,Miert2017b}. We show that the spin magnetoelectric polarizability (i.e., the polarization response to a Zeeman field) of these 1D antiferromagnetic insulators carries a signature of this quantized fractional polarization, and hence of the topology. The combined study of 1D and 2D models therefore extends our understanding of how topology determines the magnetoelectric properties of antiferromagnetic insulators. 

The remainder of this paper is organized as follows. In Sec.~\ref{sec:general} we introduce the microscopic formulation of the magnetoelectric response of Dirac antiferromagnets establishing the central result of our study. Sec.~\ref{sec:square} examines the quasitopological magnetoelectric response of the spin-orbit coupled N\'eel antiferromagnet in the buckled square lattice. In Sec.~\ref{sec:noncollinear} we prove that the quasitopological magnetoelectric response can appear even in the complete absence of spin-orbit coupling in a model for a non-collinear square lattice antiferromagnet. We examine topological signatures in the magnetoelectric response of one-dimensional antiferromagnets in Sec.~\ref{sec:1D}. Finally, we present our conclusions in Sec.~\ref{sec:conclusions}.

\section{Spin multipolization and magnetoelectric polarizability of 2D Dirac antiferromagnets \label{sec:general}}

In this section, we begin with a brief introduction to the general microscopic formulation of spin multipolization and spin magnetoelectric polarizability. We then apply this general formulation to a broad class of four-band models and obtain simple appealing expressions for the polarizability and multipolization. The latter are employed to establish and interpret the central result of this work: the magnetoelectric response of Dirac antiferromagnets in two dimensions and its 
topological origin. 

\subsection{Spin multipolization and magnetoelectric polarizability of bulk systems \label{ssec:multipolization}}

The starting point for a microscopic calculation of the spin multipolization is the electronic energy band structure of a periodic crystal, as defined by the energies $\varepsilon_{\bk n}$ and the cell-periodic part of the Bloch wave function $| u_{\bk n} \rangle$. In what follows we will often suppress momentum dependence and simply denote these $\varepsilon_{ n}$ and $| n \rangle$, respectively. Following Ref.~\onlinecite{Gao:2018p134423}, the spin multipolization $O_{ij}$ is then given by 
\be
O_{ij} = g\mu_B  \int \frac{d^d\bk}{(2\pi)^d}  \sum_{n,m} (2\mu-\varepsilon_{n}- \varepsilon_{m} ) \frac{\text{Im} [  \hat v_i^{nm }\hat s_j ^{mn}]}{(\varepsilon_{n}- \varepsilon_{m})^2}, \label{eq:multipolization}
\ee
where $\hat v_i$ is the velocity operator given by $ \hbar \hat v_i^{nm } = \partial_i H_\bk$, and $\hat s_j$ is the spin operator. Their matrix elements in the band basis are denoted $ \hat v_i^{nm } = \langle n | \hat v_i | m \rangle$ and $ \hat s_j^{mn } = \langle m | \hat s_j | n \rangle$, respectively. The sum in Eq.~\eqref{eq:multipolization} should be understood as
\be 
 \sum_{n,m} \equiv \sum_{n \in \text{occ} }  \sum_{m \in \text{unocc} }, \label{eq:band-sums}
\ee
which is to say that $n$ labels occupied bands and $m$ labels unoccupied bands. (We adhere to this convention in what follows.) Note further that in Eq.~\eqref{eq:multipolization} $g$ is the gyromagnetic factor, $\mu_B$ is the Bohr magneton, $\mu$ is the chemical potential, and $d$ is the spatial dimensions, which in this work is either $d=2$ or $d=1$.
 
It is natural to decompose the spin multipolization $O_{ij}$ into irreducible components of distinct symmetry. The monopolization $M$, for instance, is given by the scalar $M = \sum_i O_{ii}/3$, and the toroidization is given by
\be
T_i  = \frac12 \epsilon_{ijk}O_{jk} \label{eq:toroidization}.
\ee
The quadrupolization is given by the remaining five symmetric and traceless components 
\be
Q_{ij} = \frac12 (O_{ij} +O_{ji} - 2 M \delta_{ij} ). \label{eq:quadrupolization}
\ee
These quantities are the microscopic densities associated with a magnetoelectric monopole moment, magnetic toroidal dipole moment, and magnetic quadrupole moment. 

As pointed out in Ref.~\onlinecite{Gao:2018p134423}, in the case of insulators the spin multipolization is intimately related to the spin magnetoelectric polarizability (SMP), which describes the polarization response to a Zeeman magnetic field. Specifically, the SMP is obtained from the multipolization via
\be
\alpha_{ij} = \left. \frac{\partial P_i}{\partial B_j}\right|_{B \rightarrow 0} = - e \frac{\partial O_{ij}}{\partial \mu}. \label{eq:alpha_ij-M_ij}
\ee
It then follows that the microscopic expression for the SMP is given by
\be
\alpha_{ij} =-eg\mu_B \int \frac{d^d\bk}{(2\pi)^d}  \sum_{n,m}   \frac{ 2\text{Im} [  \hat v_i^{nm }\hat s_j ^{mn}]}{(\varepsilon_{n}- \varepsilon_{m})^2}. \label{eq:SMP}
\ee
This expression for the SMP is the main focus of our work. We will use Eq.~\eqref{eq:SMP} to compute $\alpha_{ij} $ for models describing classes of magnets in two dimensions, as well as simplified toy models in one dimension.

The spin magnetoelectric polarizability may be compared to the \emph{orbital} magnetoelectric polarizability studied in detail in Ref.~\onlinecite{Essin:2010p205104}. Whereas the former originates from the Zeeman coupling of the magnetic field to the electron spin, the latter derives from the orbital motion of electrons in a magnetic field. In Appendix~\ref{app:units} we briefly comment on the relation between the two, in particular as far as the units are concerned.

As is the case for the orbital magnetoelectric polarizability~\cite{Essin:2010p205104,Malashevich:2010p053032}, a Maxwell relation of the form
\be
 \frac{\partial P_i}{\partial B_j} =  \frac{\partial M_j}{\partial E_i}, \label{eq:maxwell}
\ee
holds for the SMP~\cite{Xiao:2022p086602,Xiao:2023p166302}. Importantly, since the left hand side describes a polarization response to a Zeeman field, the right hand side describes a \emph{spin} magnetization response to an electric field. Here the spin magnetization is defined as
\be
M_i = \langle \hat m_i \rangle =  \frac{g \mu_B}{\hbar} \langle \hat s_i \rangle, \label{eq:M_i-s_i}
\ee
where $\langle \hat s_i \rangle$ is the expectation value of spin. 

\subsection{Magnetoelectric polarizability of four-band models \label{ssec:tight-binding}}

Before turning to Dirac semimetals in 2D---our primary object of interest---we apply the formula for the SMP given by Eq.~\eqref{eq:SMP} to the general class of four-band models. 
Two
models studied in this work fall in this class. Furthermore, as we will now show, simplified expressions for the SMP can be obtained in the case of four-band models, which do not require determining the eigenstates. 

Consider a class of four-band models defined by the Hamiltonian
\be
H = \sum_{\bk} c^\dagger_\bk H_\bk c_\bk, \;\; c_{\bk} = ( c_{\bk A\up},c_{\bk A\down},c_{\bk B\up},c_{\bk B\down})^T, \label{eq:H-def}
\ee
where $c_{\bk \alpha \sigma }$ are electron destruction operators with sublattice index $\alpha=A,B$ and spin $\sigma = \up,\down$. Note that here we refer to $A$ and $B$ simply as sublattices, but they should be regarded as a general internal pseudospin degree of freedom (e.g.~sublattice, orbital, valley). In this basis, the Hamiltonian matrix $H_\bk $ can be written as
\be
H_\bk = f_x \tau^x+ f_y \tau^y+  \tau^z \bn \cdot \bsigma, \label{eq:H_k}
\ee
where $\btau = (\tau^x,\tau^y,\tau^z)$ and $\bsigma=(\sigma^x,\sigma^y,\sigma^z)$ are two sets of Pauli matrices corresponding to the sublattice and spin degrees of freedom, respectively, and $\bn = (n_x, n_y,n_z)$ is a three-component vector. The coefficients $f_{x,y}$ and $\bn$ are in general momentum-dependent functions. Assuming that spatial inversion symmetry ($\mathcal I$) exchanges the sublattices and that time-reversal ($\mathcal T$) acts in the standard way for spin-$1/2$, the Hamiltonian $H_\bk$ takes the most general form compatible with the product of $\mathcal I \mathcal T$. The latter symmetry is important for our purpose, since our goal is to consider antiferromagnets which allow for a magnetoelectric polarizability. Imposing either $\mathcal I $ or $\mathcal T$ on $H_\bk$ in Eq.~\eqref{eq:H_k} would forbid a nonzero $\alpha_{ij}$. 

It is straightforward to find the energy spectrum of $H_\bk$. It consists of two manifestly twofold degenerate spectral branches $\varepsilon_{1}$ and $\varepsilon_{2}$ (i.e., a valence and a conduction band), which are given by 
\be
\varepsilon_{1,2} = \mp  \varepsilon_\bk, \quad \varepsilon_\bk = \sqrt{f_x ^2 + f_y^2 + \bn^2 }.
\ee
Note that the twofold degeneracy is a consequence of the product of $\mathcal I$ and $\mathcal T$ symmetry. The projectors onto the valence ($P_-$) and conduction ($P_+$) bands take the simple form
\be
P_\pm = \frac12(\mathbb{1}\pm H_\bk/\varepsilon_\bk), \label{eq:projectors}
\ee
and it is only these projectors which are needed to compute the SMP $\alpha_{ij}$. This follows from the fact that for $H_\bk$ given by Eq.~\eqref{eq:H_k}, the integrand of Eq.~\eqref{eq:SMP} can be expressed as
\be
\sum_{n,m}   \frac{ \text{Im} [  \hat v_i^{nm }\hat s_j ^{mn}]}{(\varepsilon_{n}- \varepsilon_{m})^2}= \frac{1}{4\varepsilon_\bk^2} \text{Im}  \text{Tr}[ P_- \hat v_i P_+ \hat s_j ] , \label{eq:integrand-projector}
\ee
where we recall Eq.~\eqref{eq:band-sums}. The trace is easily taken and yields
\be
 \text{Im}  \text{Tr}[ P_- \hat v_i P_+ \hat s_j ] = \frac{\partial_i \bn  \times \bn\cdot \ehat_j}{ \varepsilon_\bk},   \label{eq:Im-Tr-4band}
\ee
leading to an appealingly simple formula for the SMP given by
\be
\alpha_{ij} =  eg\mu_B \int \frac{d^d\bk}{(2\pi)^d} \frac{\partial_i\bn    \times \ehat_j \cdot \bn }{2 \varepsilon_\bk^3}. \label{eq:SMP-4band}
\ee
For the class of Hamiltonians considered here, the expression for the SMP can thus be recast entirely in terms of the energy dispersion $\varepsilon_\bk$ and the Hamiltonian (via $\bn$). This makes it particularly straightforward to compute the SMP, as an explicit calculation of the eigenstates is not required. 

Based on the calculation of $\alpha_{ij} $, it is straightforward to obtain a similar expression for the multipolization $O_{ij}$ given by Eq.~\eqref{eq:multipolization}. While the most general expression of $\alpha_{ij}$ did not require specifying a uniform dispersion $\varepsilon_{0,\bk} $ in $H_\bk$ of Eq.~\ref{eq:H_k} (i.e., a term proportional to the identity), such a term does enter $O_{ij}$. Therefore, if a uniform dispersion $\varepsilon_{0,\bk} $ is included in $H_\bk$, the multipolization of a general four-band model reads as
\be
O_{ij} =  g\mu_B  \int \frac{d^d\bk}{(2\pi)^d} (\varepsilon_{0,\bk} - \mu ) \frac{\partial_i\bn    \times \ehat_j \cdot \bn }{2 \varepsilon_\bk^3}, \label{eq:Oij-4band}
\ee
from which Eq.~\eqref{eq:SMP-4band} is recovered via Eq.~\eqref{eq:alpha_ij-M_ij}.

The expressions for the SMP and the multipolization given by Eqs.~\eqref{eq:SMP-4band} and \eqref{eq:Oij-4band}, respectively, have been obtained for the class of four-band models given by Eq.~\eqref{eq:H_k}. This formulation is natural when the microscopic degrees of freedom are the sublattice (or layer, or other generalized orbital degree of freedom) and spin. A more general formulation of four-band models which does not make reference to any specific representation is discussed in Appendix~\ref{app:four-band}. We will make use of the obtained generalization of Eq.~\eqref{eq:SMP-4band} in Sec.~\ref{sec:noncollinear}.

\subsection{Magnetoelectric response of antiferromagnetic Dirac semimetals in two dimensions \label{ssec:2D-dirac}}

\begin{figure}
	\includegraphics[width=0.95\columnwidth]{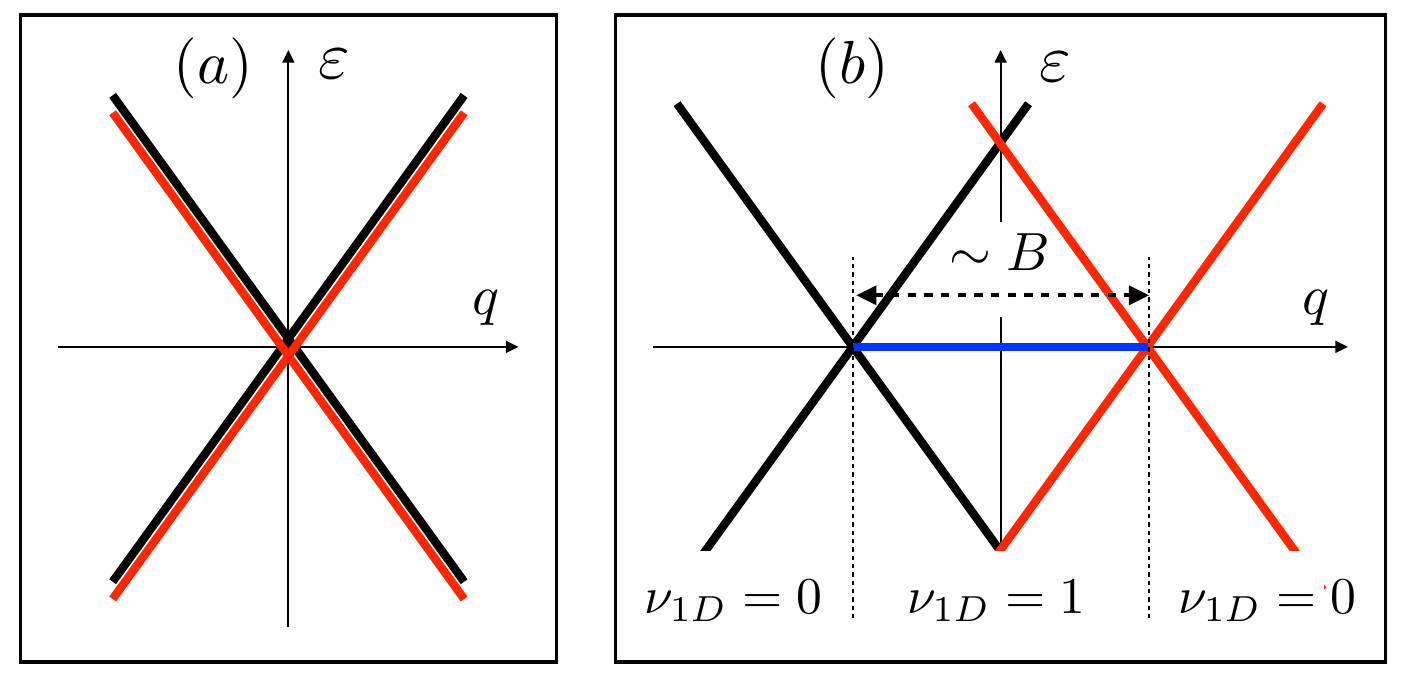}
	\caption{(a) Schematic of the linear dispersion of a fourfold degenerate Dirac crossing at $\bq=0$. The combination of inversion and time-reversal symmetry ensure a twofold degenerate spectrum away from the Dirac crossing. (b) Schematic of the splitting of the Dirac point into (2D) Weyl points in the presence of a Zeeman field $B$. The momentum separation $\delta q$ is proportional to $B$ and creates a region of effectively one-dimensional (1D) topologically nontrivial systems in the direction perpendicular to the separation. }
	\label{fig:dirac_sketch}
\end{figure}

Given these simplified expressions for the spin multipolization and magnetoelectric polarizability, we now consider the application of these formulae to Dirac semimetals in two dimensions. 

Consider a fourfold degenerate Dirac point in two dimensions described by the Hamiltonian 
\be
\mathcal H_\bq = \hbar v   \tau^z  \bq\times \bsigma , \label{eq:H-dirac}
\ee
where $\bq=(q_x,q_y)$ is the wave vector and $v$ is the Fermi velocity. This continuum Dirac Hamiltonian falls in the general class of Hamiltonians given by Eq.~\eqref{eq:H_k}. The Pauli matrices $\bsigma$ correspond to spin and the cross product should be understood as $\bq\times \bsigma = \epsilon_{ij} q_i \sigma^j$, where $\epsilon^{ij}$ is the two-dimensional Levi-Civita tensor. In the context of this Dirac Hamiltonian the Pauli matrices $\btau$ correspond to a valley or flavor index, which is required by fermion doubling. (Dirac fermions in two dimensions must come in pairs~\cite{Semenoff:1984p2449}.) The linear energy spectrum described by Eq.~\eqref{eq:H-dirac} is sketched in Fig.~\ref{fig:dirac_sketch}(a), where red and black indicate the two flavors.

Since the Dirac Hamiltonian of Eq.~\eqref{eq:H-dirac} describes a gapless Dirac theory, it is predicated on the tacit assumption that a mass term is forbidden by symmetry. Indeed, fourfold degenerate band crossings with linear dispersion require the presence of additional crystal symmetries beyond $\mathcal I$ and $\mathcal T$ symmetry~\cite{Yang:2014p4898,Young:2015p126803}. (A specific example will be discussed in Sec.~\ref{sec:square}.) Here we simply assume such symmetries are present in the nonmagnetic state and focus on the symmetry breaking mass term introduced by antiferromagnetic N\'eel order, in particular N\'eel order along the $z$ direction. In the presence of antiferromagnetic N\'eel order, the Dirac Hamiltonian takes the form
\be
\mathcal H_\bq = \hbar v   \tau^z  \bq\times \bsigma + N_z \tau^z \sigma^z ,  \label{eq:H_N-dirac}
\ee
where $N_z$ is the $z$-component of the N\'eel vector. It is clear that $N_z$ enters as a mass term and that the Hamiltonian describes a gapped Dirac antiferromagnet. Consider the insulator obtained from a filled valence band. Since N\'eel order breaks both $\mathcal I$ and $\mathcal T$, but preserves $\mathcal I \mathcal T$ symmetry, this insulator in principle allows for nonzero $\alpha_{ij}$. In-plane rotation symmetry furthermore dictates that $\alpha_{xx} =\alpha_{yy} $ and that the off-diagonal components vanish. 

Our goal is now to compute the magnetoelectric polarizability $\alpha_{ij}$ for the antiferromagnetic Dirac insulator. To compute $\alpha_{xx} $, we use Eq.~\eqref{eq:SMP-4band} and find
\begin{align}
\alpha_{xx}  &= -\frac12 eg\mu_B \int \frac{d^2\bq}{(2\pi)^2} \frac{\hbar v N_z }{(\hbar^2v^2q^2+N_z^2)^{3/2}}, \label{eq:a_xx-Dirac} \\
& = -\frac{ eg\mu_B}{4\pi \hbar v} \text{sgn}(N_z).
\end{align}
The full polarizability tensor immediately follows, and can be written as
\be
\alpha_{ij} = - -\frac{ eg\mu_B}{4\pi \hbar v}\text{sgn}(N_z) \delta_{ij}. \label{eq:alpha_ij-Dirac2D} %
\ee
This expression for the spin magnetoelectric polarizability of a Dirac antiferromagnet in 2D constitutes the central result of this work. Its significance derives from its topological origin, which is evidenced by the dependence on the sign of $N_z$ only. This dependence appears counterintuitive, since the symmetries which forbid a magnetoelectric response, i.e., inversion ($\mathcal I$) and time-reversal symmetry ($\mathcal T$), are restored as $N_z \rightarrow 0 $, yet Eq.~\eqref{eq:alpha_ij-Dirac2D} retains its $\mathcal I$-odd and $\mathcal T$-odd contribution with an ambiguous sign. This behavior is reminiscent of the parity anomaly in ($2+1$)-dimensional QED, which is known to have a topological origin~\cite{Affleck:1982p413,Niemi:1983p2077,Redlich:1984p18,Jackiw:1984p2375}. 

Two observations regarding the role of topology can be made. First, up to a factor of $\hbar v$ the integrand of \eqref{eq:a_xx-Dirac} is equal to the Berry curvature of a gapped Dirac fermion in 2D~\footnote{In this case Dirac fermion refers to a twofold Dirac crossing in two dimensions, i.e., a linear crossing of two nondegenerate bands. Such Dirac points must always come in pairs, which is why Eq.~\eqref{eq:H-dirac} describes two such Dirac crossings, together forming a fourfold Dirac point.}, thus suggesting that $\alpha_{ij}$ captures a topological property of 2D Dirac fermions. It is important to emphasize, however, that Eq.~\eqref{eq:alpha_ij-Dirac2D} is not related to a strong topological index in 2D, such as the Chern number. Indeed, since the two flavors of Dirac fermions described by Eq.~\eqref{eq:H-dirac} have opposite Berry curvature when gapped by $\mathcal H_N$, the Chern number of the antiferromagnetic insulator vanishes. 

The second observation is that the magnetoelectric response of Eq.~\eqref{eq:alpha_ij-Dirac2D} is not simply determined by fundamental constants, but depends on microscopic details via the velocity $v$. The velocity may be exchanged for a length scale of the order of the lattice constant by  defining $v \equiv \hbar/ m_e a$, such that the prefactor $eg\mu_B/4\pi \hbar v$ is recast as $ g a e^2/4h$. This further emphasizes the dependence on microscopic details (via $a$) and is another indicator that the magnetoelectric response does not express a genuine two-dimensional topological invariant.

To obtain a more precise understanding of the topological origin of Eq.~\eqref{eq:alpha_ij-Dirac2D}, let us re-examine the electromagnetic response of topological semimetals in 2D.

\subsection{Polarization and quasitopological electromagnetic response of Dirac semimetals \label{ssec:em-dirac}}

The general structure of electromagnetic responses in topological semimetals was studied in detail by Ramamurthy and Hughes~\cite{Ramamurthy:2015p085105}. They demonstrated that Dirac semimetals in 2D have a nontrivial electric polarization, which is proportional to the Dirac node separation in wave vector space. Specifically, the polarization $P_i$  is given by ($i,j=x,y$)
\be
P_i  = \frac{e}{2\pi} \epsilon^{ij}b_j \label{eq:P_i-a_j},
\ee
where $2\bb$ is the wave vector space separation between the Dirac nodes. The origin of this polarization can be understood by viewing a Dirac semimetal in 2D---more precisely a Dirac semimetal of the kind studied in Ref.~\onlinecite{Ramamurthy:2015p085105}---as a phase intermediate between two topologically distinct 2D insulators: a trivial insulator with polarization $P_i =ne\nu_i/a$ and an insulator with polarization $P_i =(n+1/2)e\nu_i/a$, where $n$ is an integer and $a$ is the lattice constant. The latter can simply be viewed as a stack of 1D crystalline insulators with a nontrivial $\mathbb{Z}_2$ index and half-integer quantized polarization $e/2~\textrm{mod}~e$~\cite{Hughes:2011p245132}. The 2D Dirac semimetal has a polarization proportional to the separation of the Dirac nodes, thus interpolating between the two insulators, since the distance between the nodes (equal to $2\bb$) marks a region of effectively 1D Hamiltonians with half-integer polarization (in a direction perpendicular to $\bb$). This is shown schematically in Fig.~\ref{fig:dirac_sketch}(b). It then follows that the polarization described by Eq.~\eqref{eq:P_i-a_j} is ultimately a consequence of a $\mathbb{Z}_2$ topological invariant
in 1D, i.e., a dimensionally reduced index. For this reason, the polarization of a Dirac semimetal is more appropriately referred to as a quasitopological electromagnetic response~\cite{Ramamurthy:2015p085105}. 

To establish a connection between Eq.~\eqref{eq:alpha_ij-Dirac2D} and Eq.~\eqref{eq:P_i-a_j}, consider the Dirac Hamiltonian of Eq.~\eqref{eq:H-dirac} in the presence of the Zeeman field $\bB$ given by 
\be
H_{B} = - \frac12 g \mu_B \bB\cdot \bsigma.  \label{eq:H_B-dirac}
\ee
The full Hamiltonian $\mathcal H_\bq + \mathcal  H_{B} $ can be written as
\be
\mathcal H_\bq + \mathcal  H_{B} = \hbar v \tau^z (\bq - \bb\tau^z)\times \bsigma, \label{eq:H_q-H_B}
\ee
with $\bb$ given by
\be
(b_x, b_y) = \frac{g \mu_B}{2\hbar v } (B_y, -B_x). \label{eq:a_x-a_y}
\ee
It follows that in the presence of a Zeeman field the Dirac nodes remain gapless, but are separated in momentum space by a distance $2\bb$. The separation vector $\bb$ can be written as $b_i = (g \mu_B/2\hbar v)\epsilon^{ij}B_j  $, such that the polarization given by the general formula Eq.~\eqref{eq:P_i-a_j} becomes $P_i  = - (e g \mu_B/4\pi\hbar v)\delta_{ij}B_j$. Taking the derivative $\partial P_i /\partial B_j$ yields $\alpha_{ij} = - (e g \mu_B/4\pi\hbar v)\delta_{ij}$, which exactly equals the result of Eq.~\eqref{eq:alpha_ij-Dirac2D}, except for the factor $\text{sgn}(N_z)$. The latter arises from a subtlety left implicit in Eq.~\eqref{eq:P_i-a_j}. To determine this sign of the quantized polarization of an insulator in 1D (i.e., whether it is $+e/2$ or $-e/2$), and by extension the sign of $P_x$ in \eqref{eq:P_i-a_j},  it is necessary to introduce a regulating symmetry breaking mass term, which leaves a sign trace when taken to zero at the end of the calculation~\cite{Qi:2008p195424,Ramamurthy:2015p085105}. In our case, $N_z$ represents the symmetry breaking mass---indeed, $N_z$ breaks $\mathcal I$ and $\mathcal T$---and thus fixes the sign of $\alpha_{ij}$. 

This analysis demonstrates that the polarization response of the Dirac antiferromagnet to a Zeeman field is ultimately a manifestation of the more general electromagnetic response theory of topological Dirac semimetals in 2D, and thus exposes its topological origin. It is important to point out a key difference with the response theory derived by Ramamurthy and Hughes in Ref.~\onlinecite{Ramamurthy:2015p085105}, however. The theory considered in Ref.~\onlinecite{Ramamurthy:2015p085105} is based on a model for a Dirac semimetal without spin-orbit coupling (i.e., a spinless model), and in that model the wave vector $\bb$ is odd under inversion ($\mathcal I$) and \emph{even} under time-reversal ($\mathcal T$). In contrast, in the model considered here, which describes a spin-orbit coupled Dirac semimetal, $\bb$ is proportional to the Zeeman field, and therefore odd under time-reversal. As a result, in the present context the polarization response constitutes a fundamentally different effect, namely a magnetoelectric effect. The polarization (odd under $\mathcal I$) is linearly proportional to the Zeeman field (odd under $\mathcal T$) and the proportionality constant depends on the sign of $N_z$, which is indeed odd under both $\mathcal I$ and $\mathcal T$ but even under $\mathcal I \mathcal T$. Hence, in antiferromagnets the 
quasitopological electromagnetic response of Dirac semimetals can manifest as a quasitopological magnetoelectric effect.

\subsection{Summary of main result \label{ssec:summary}}

Before proceeding to a demonstration of the quasitopological magnetoelectic effect in microscopic lattice models, we briefly summarize the main result of this section and discuss its general validity.

We have shown that the spin magnetoelectric polarizability of a special class of insulating antiferromagnets, which we refer to as Dirac antiferromagnets, contains a contribution
originating from the topology of the ground state.
This result was obtained by considering a continuum Dirac model coupled to antiferromagnetic N\'eel order, as given by Eq.~\eqref{eq:H_N-dirac}. Such model describes a spin-orbit coupled insulator with collinear antiferromagnet N\'eel order, and it is the $\mathcal I \mathcal T$-symmetric N\'eel order parameter which provides a gap for the gapless Dirac states. Crucially, the 
topological 
contribution to the spin magnetoelectric polarizability is not limited to Dirac antiferromagnets of this particular kind, but can arise in a broader class of antiferromagnets. The key requirement is that the antiferromagnetic insulator is properly described by an effective Dirac theory. This will be demonstrated in detail in Sec.~\ref{sec:noncollinear}, where we analyze a minimal model of a noncollinear antiferromagnet without spin-orbit coupling. Within the framework of this model, the Dirac dispersion arises from the noncollinear magnetic state itself and the gap is produced by non-magnetic bond density wave order. It is therefore important to emphasize that our main result applies to a broad class of antiferromagnets beyond spin-orbit coupled collinear N\'eel states. 

A second remark concerns the relation between the spin magnetoelectric polarizability and the spin multipolization. Although we have primarily focused on the magnetoelectric polarizability in Secs.~\ref{ssec:2D-dirac} and \ref{ssec:em-dirac}, immediate implications for the spin multipolization follow from Eq.~\eqref{eq:Oij-4band}. It follows in particular that Dirac antiferromagnets have a
quasitopological spin multipolization given by
\be
O_{ij} =  ( \mu - \varepsilon_D)\frac{g \mu_B}{4 \pi \hbar v}\text{sgn}(N_z) \delta_{ij}, \label{eq:O_ij-Dirac2D}
\ee
where $\varepsilon_D$ is the energy at which the Dirac crossing occurs. Clearly, for a rotationally symmetric Dirac dispersion the only nonzero component is the monopolization.
In general, when symmetries are lowered (see Secs.~\ref{sec:square} and \ref{sec:noncollinear}), the toroidization and quadrupolization may be nonzero.

\section{Spin-orbit coupled square lattice antiferromagnet \label{sec:square}}

\begin{figure}
	\includegraphics[width=0.95\columnwidth]{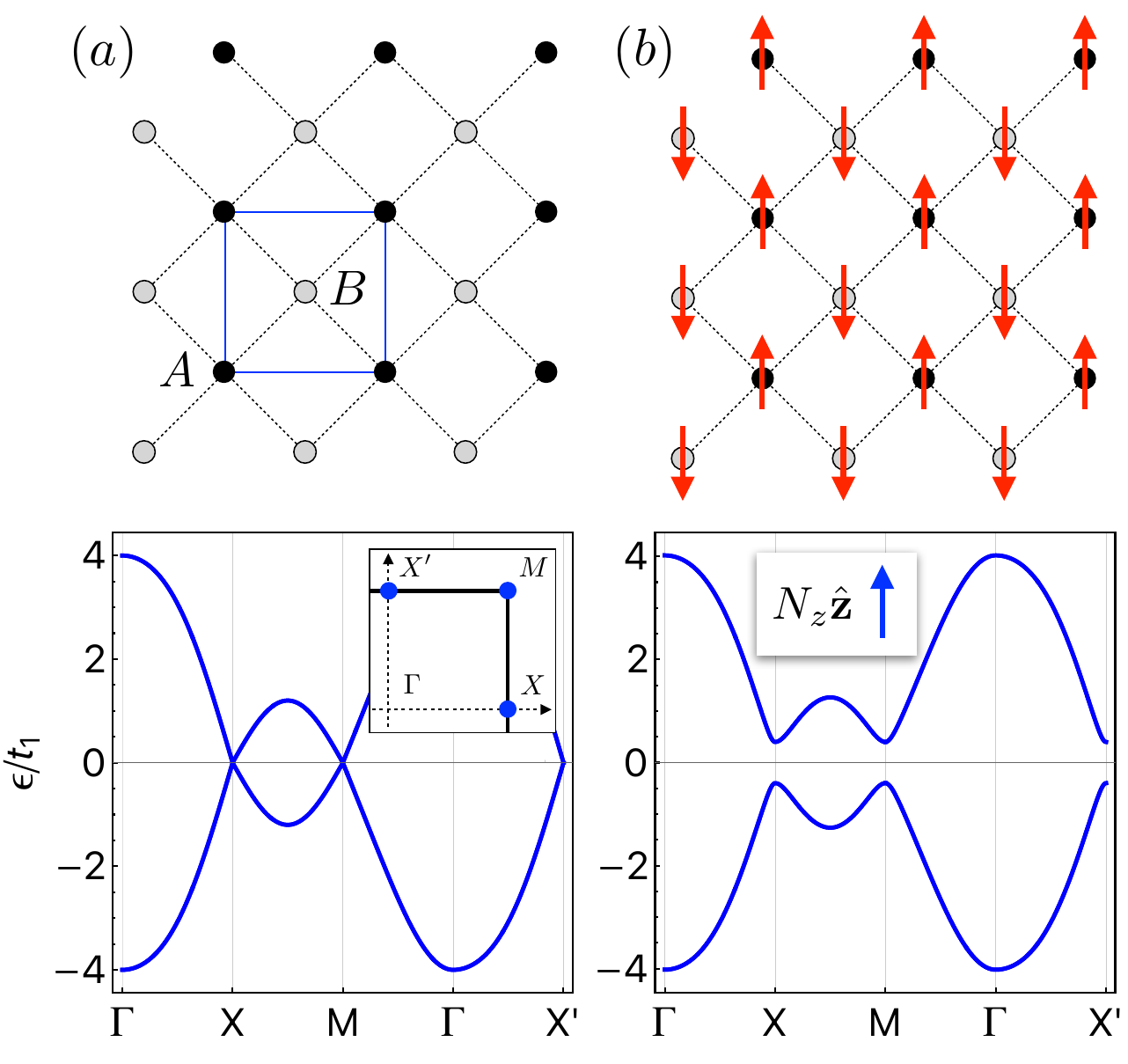}
	\caption{(a) Top panel: Sketch of buckled square lattice in two dimensions. The $A$ and $B$ (black and gray dots) sites a shifted in the negative and positive $z$ direction (i.e., out of the plane), respectively. Bottom panel: The spectrum of Eq.~\eqref{eq:H_k-square}, shown for $\lambda=0.6t_1$, featuring Dirac points at $M$, $X$, and $X'$ (see inset). (b) Top panel: Sketch of inversion-breaking N\'eel order in the buckled square lattice model. When the N\'eel vector points in the $z$ direction all three Dirac points are gapped (bottom panel).}
	\label{fig:buckled}
\end{figure}

We now turn to a detailed analysis of magnetoelectric coupling in a microscopic model for spin-orbit coupled antiferromagnets. The model in question is defined on a buckled square lattice, as shown in Fig.~\ref{fig:buckled}, and was first introduced in Ref.~\onlinecite{Young:2015p126803} to showcase the existence of symmetry-enforced Dirac points in 2D. Here we consider the 
additional
presence of antiferromagnetic N\'eel order, described by a N\'eel vector $\bN$, which breaks both inversion and time-reversal symmetry but preserves their product~\cite{Wang:2017p115138,Smejkal:2017p106402}. The N\'eel antiferromagnet therefore allows for nonzero $\alpha_{ij}$. 

In what follows we first examine the symmetries of the buckled lattice antiferromagnet and determine the allowed components of the SMP for a given direction of $\bN$. To make a connection with the general analysis of Sec.~\ref{sec:general}, we then consider the continuum Dirac theory of the buckled lattice, obtained by expanding the tight-binding lattice Hamiltonian around the high-symmetry points. This is where the symmetry-enforced Dirac points are located. We then compute and discuss the magnetoelectric polarizability based on Eqs.~\eqref{eq:SMP} and \eqref{eq:SMP-4band}. 

\subsection{Model and symmetries \label{ssec:square-model}}

The square lattice model shown in Fig.~\ref{fig:buckled}(a) consists of two square sublattices, denoted $A$ and $B$, which are displaced in the $z$ direction, i.e., the direction perpendicular to the lattice plane.  As a result of this vertical displacement (``buckling''), the space group of the crystal lattice is $P4/nmm$, and a minimal generic tight-binding Hamiltonian compatible with the space group symmetries takes the form~\cite{Young:2015p126803}
\be
H_{0,\bk} = -4 t_1 c_{x/2}c_{y/2}  \tau^x  - 2\lambda  \tau^z(s_x \sigma^y - s_y \sigma^x ), \label{eq:H_k-square}
\ee
where we have used the description of the electronic degrees of freedom introduced in Eq.~\eqref{eq:H-def}. We have further abbreviated $c_{i/2} \equiv \cos (k_i/2)$ and $s_{i} \equiv \sin k_i$. (We set the lattice constant $a=1$.) In terms of the general form of $H_\bk$ defined in Eq.~\eqref{eq:H_k}, we have $f_x = -4 t_1 c_{x/2}c_{y/2}$ and $(n_x,n_y) = 2\lambda (s_y,-s_x)$. Here $t_1$ denotes the nearest neighbor hopping and $\lambda $ is a spin-orbit coupling energy scale. In the Hamiltonian of Eq.~\eqref{eq:H_k-square} we have neglected an allowed uniform dispersion $\varepsilon_{0,\bk} $, which is given by $\varepsilon_{0,\bk} =  -2 t_2(c_x+c_y)$ when second-nearest neighbor hopping is included, since it does not affect the key results. The magnetic part of the Hamiltonian takes the form
\be
H_{N,B} =  \tau^z \bN \cdot \bsigma - \frac12 g \mu_B \bB\cdot \bsigma , \label{eq:H_N,B}
\ee
where the first term describes N\'eel order and the second term accounts for a Zeeman field (see above). The full Hamiltonian then becomes $H_\bk = H_{0,\bk} + H_{N,B}$. 

As mentioned, the nonmagnetic buckled square lattice has space group $P4/nmm$, which is generated by the fourfold rotation $ \{ \mathcal C_{4z} | \tfrac12 0 \} $, the twofold screw rotation $  \{ \mathcal C_{2x} | \tfrac12 0 \}$, and the inversion $\{ \mathcal I  | 0 0 \}$. The algebraic structure of the space group $P4/nmm$ dictates that all energy bands at points of high symmetry on the Brillouin zone boundary must be fourfold degenerate, thus giving rise to stable Dirac points at the $M$, $X$, and $X'$ points indicated in the inset of Fig.~\ref{fig:buckled}(a)~\cite{Young:2015p126803}. The algebraic space group relations and their implications for band degeneracies are discussed in more detail in Appendix~\ref{app:P4/nmm}. Lowering the space group symmetry affects the stability of the Dirac points and may gap one or all of the Dirac crossings depending on the pattern of symmetry breaking.   

AFM N\'eel order is an example of such symmetry breaking, with different orientations of the N\'eel vector $\bN$ leading to distinct patterns of symmetry breaking. As mentioned in the beginning of this section, inversion symmetry is broken for any direction of the N\'eel vector, as is time-reversal symmetry. When the N\'eel vector points in the $z$ direction, all Dirac points acquire a gap, as is shown in Fig.~\ref{fig:buckled}(b). Instead, when $\bN$ points in the $x$ direction, the glide mirror symmetry $\{\mathcal M_{x} | \tfrac120 \}$ is preserved, with the consequence that the fourfold crossings on the $k_x = \pi$ line remain protected but are no longer pinned to the high symmetry points $M$ and $X$~\cite{Smejkal:2017p106402,Wang:2017p115138}. (A similar result obviously holds when the N\'eel vector is in $y$ direction.) For the purpose of calculating the SMP in Sec.~\ref{ssec:SMP-buckled} we will choose the N\'eel vector in the $z$ direction.

\begin{figure}
	\includegraphics[width=0.95\columnwidth]{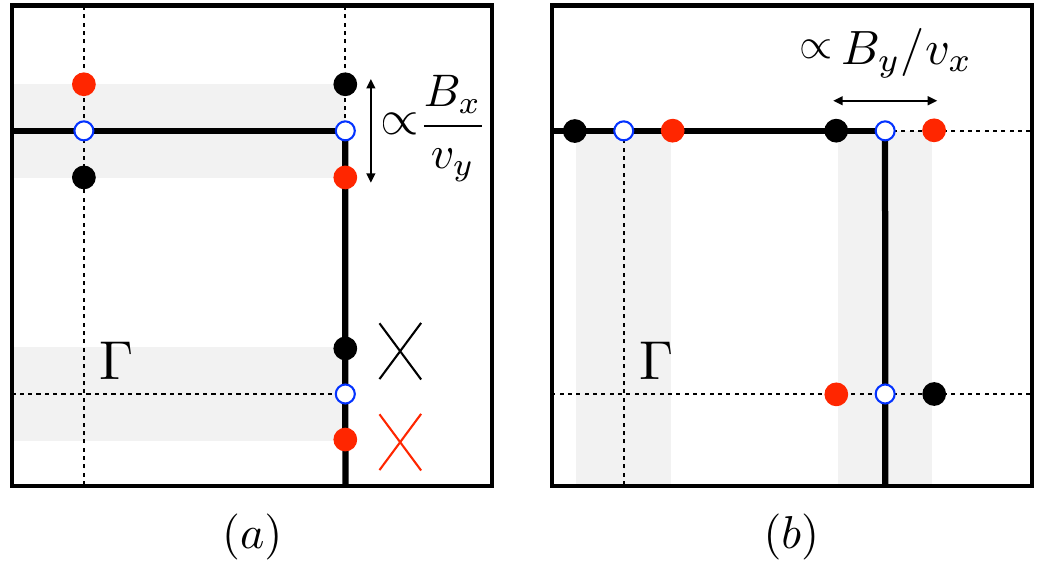}
	\caption{Splitting of the fourfold Dirac points in the presence of a Zeeman field. (a) A Zeeman field $B_x$ splits the Dirac points into pairs of twofold Dirac crossings along the $k_y$ axis. The gray shadings indicate regions where the Hamiltonian, when viewed as a collection of 1D models parametrized by $k_x$, has a nontrivial $\mathbb{Z}_2$ invariant with half-quantized polarization. (b) Similarly, a Zeeman field $B_y$ splits the Dirac points along the $k_x$ axis, creating a family of nontrivial 1D insulators along $k_x$. }
	\label{fig:buckled_zeeman}
\end{figure}

\subsection{Landau theory of magnetoelectric couplings \label{ssec:square-landau}}

As a next step, we determine the allowed magnetoelectric couplings in the N\'eel-ordered state from a general symmetry perspective. An insightful approach to understanding the magnetoelectric properties of N\'eel order is to consider the allowed trilinear couplings between N\'eel vector $\bN$, electric polarization $\bP$, and magnetization $\bM$, within the framework of Landau theory. Specifically, we seek to determine the allowed trilinear couplings of the general form $\mathcal F_{P,M,N} =  \beta_{ijk} P_i M_j N_k$. This is achieved by examining how the components of these three fields transform under space group symmetries and then constructing all possible invariants. The reason such couplings are in principle allowed is the fact that N\'eel order breaks $\mathcal I$ and $\mathcal T$, but preserves the product $\mathcal I \mathcal T$. In general, such couplings are forbidden in a conventional N\'eel antiferromagnet, which breaks translation symmetry and retains inversion centers. Appendix~\ref{app:landau} elaborates on the description of magnetoelectric effects based on trilinear couplings in Landau theory. 

We find that the allowed couplings in the case of the square lattice model are given by
\begin{multline}
\mathcal F_{P,M,N} =  \beta_{zzz} P_z M_z N_z + \beta_{xxz}(P_x M_x + P_y M_y )N_z \\
+ \beta_{zxx} ( P_zM_xN_x +P_zM_yN_y)   \\
+\beta_{xzx} ( P_xM_zN_x +P_yM_zN_y)  . \label{eq:trilinear}
\end{multline}
For given direction of the N\'eel vector, Eq.~\eqref{eq:trilinear} allows us to determine both the magnetoelectric multipole moment induced by N\'eel order and the corresponding linear magnetoelectric response. For instance, when the N\'eel vector is along the $\hat z$ axis, the magnetoelectric response is described by an effective free energy term
\be
\mathcal F _{P,M} = \bar\alpha_{zz} P_z M_z  + \bar\alpha_{xx}(P_x M_x + P_y M_y ), \label{eq:me-Nz-buckled}
\ee
where $\bar\alpha_{ii} \sim \beta_{iiz} N_z$. It follows that in this case the only nonvanishing components of the magnetoelectric tensor given by Eq.~\eqref{eq:SMP-4band} are the diagonal components. In particular, an applied magnetic (Zeeman) field in the $x$ direction will lead to an electric polarization in the same direction. (How this is established more explicitly in the context of Landau theory is discussed in Appendix~\ref{app:landau}.) It similarly follows that N\'eel order in the $\hat z$ direction corresponds to (an in-plane) magnetoelectric monopole moment. Instead, N\'eel order in the $\hat x$ (or $\hat y$) direction induces both a toroidal moment and magnetic quadrupole moment. 

This general symmetry analysis provides the basis for the microscopic analysis presented in Sec.~\eqref{ssec:SMP-buckled}.

\subsection{Fourfold Dirac fermions \label{ssec:buckled-dirac}}

To connect the analysis of the buckled square lattice to the general results presented in Sec.~\ref{sec:general}, the next step is to determine the continuum theory for the fourfold degenerate Dirac points. This is achieved by expanding the lattice Hamiltonian around the $M$ and $X$ points; the $X'$ point is related to the latter by the fourfold rotation symmetry. 

Consider first the zone corner $M$. Expanding Eq.~\eqref{eq:H_k-square} in small momentum $\bq$ around $M$ yields 
\be
\mathcal H^{M}_\bq =  \hbar v^M (q_x \tau^z\sigma^y - q_y \tau^z\sigma^x ), \quad v^M = 2\lambda/\hbar ,  \label{eq:H^M_q}
\ee
which equals the isotropic form of Eq.~\eqref{eq:H-dirac}. As a result, the N\'eel vector and a Zeeman magnetic field enter in precisely the way given by Eqs.~\eqref{eq:H_N-dirac} and~\eqref{eq:H_B-dirac}.

Consider next the $X$ point. The Hamiltonian near $X$ is expanded in small momenta $\bq$ as
\be
\mathcal H^{X}_\bq =  2t_1 q_x \tau^x  + 2\lambda q_x \tau^z \sigma^y + 2\lambda q_y \tau^z \sigma^x , \label{eq:H^X_q}
\ee
which describes a anisotropic Dirac theory with different velocities given by $v^X_y = 2\lambda/\hbar$ and 
\be
v^X_x = \frac{2}{\hbar}\sqrt{t^2_1 + \lambda^2}, \quad \sin\theta = \frac{2t_1}{\hbar v^X_x}, \; \cos\theta = \frac{2\lambda}{ \hbar v^X_x}. \label{eq:v_x-X}
\ee
Here we have introduced the angle $\theta$ to parametrize the relative contributions to $v_x$. The linearized Hamiltonian of Eq.~\eqref{eq:H^X_q} does not have the standard Dirac form (where each momentum is associated with one Dirac matrix), but can be brought into the standard form by a unitary transformation
\be
U = \cos\tfrac{\theta}{2} \mathbb{1}- i \sin\tfrac{\theta}{2} \tau^y \sigma^y = e^{- i \theta \tau^y \sigma^y/2  }.
\ee
By rotating $H^{X}_\bq$ using $U$ we obtain
\be
\mathcal H^{X}_\bq \rightarrow U^\dagger \mathcal H^{X}_\bq U =\hbar v^X_xq_x  \tau^z \sigma^y +\hbar  v^X_y q_y \tau^z \sigma^x,  \label{eq:H^X_q-rotate}
\ee
which does have the standard form, with Dirac matrices $ \tau^z \sigma^y$ and $ \tau^z \sigma^x$ associated with the two momentum components $q_x$ and $q_y$. As we will now show, bringing the Dirac Hamiltonian into this form is necessary to correctly capture the effect of the Zeeman field and use Eq.~\eqref{eq:H_q-H_B} to infer the magnetoelectric polarizability. 

To determine how N\'eel order and a Zeeman field $B_x$ enter in the continuum theory at $X$, we must perform the same rotation on $\mathcal H_{N,B}$, which yields
\begin{multline}
U^\dagger \mathcal H_{N,B}  U = \\
N_z \tau^z\sigma^z -\frac{1}{2}g\mu_B B_x (\cos\theta \sigma^x +\sin\theta \tau^y \sigma^z), \label{eq:H^X_N,B-rotate}
\end{multline}
Whereas the N\'eel order term is not affected, the Zeeman coupling is rotated. This affects the determination of the quasitopological magnetoelectric response given by Eq.~\eqref{eq:P_i-a_j}. To see this, note that, based on Eq.~\eqref{eq:H^X_N,B-rotate}, the Zeeman field enters as $\cos\theta \sigma^x +\sin\theta \tau^y \sigma^z$, and only the first term represents a shift of the Dirac point in momentum (see the discussion of Sec.~\ref{ssec:em-dirac}). Specifically, the Dirac point shifts by an amount $b_y = g\mu_B B_x \cos\theta / 2\hbar v^X_y$, which, using Eqs.~\eqref{eq:v_x-X}, reduces to $b_y = g\mu_B B_x  / 2\hbar v^X_x$. Therefore, to correctly establish $b_y$ and apply Eq.~\eqref{eq:P_i-a_j}, it is essential to bring the continuum Hamiltonian at $X$ into the form given by Eqs.~\eqref{eq:H^X_q-rotate} and \eqref{eq:H^X_N,B-rotate}. (Note that the term proportional to $\sin\theta$ in \eqref{eq:H^X_N,B-rotate} commutes with both $ \tau^z \sigma^y$ and $ \tau^z \sigma^x$, and therefore does not affect the quasitopological response.)

The effect of an applied Zeeman field on the three fourfold Dirac points at $M$, $X$, and $X'$ is summarized in Fig.~\ref{fig:buckled_zeeman}. (Note that here $N_z=0$.) Figure~\ref{fig:buckled_zeeman}(a) indicates how the Dirac points are shifted in response to $B_x$: all three fourfold Dirac points are split into twofold Dirac crossings along $k_y$. The sign of $b_y$ is determined by the direction in which the twofold Dirac points with opposite Berry phase shift, which is represented by red and black dots. The gray shadings correspond to regions where the collection of 1D Hamiltonians parametrized by $k_y$ have a nontrivial $\mathbb{Z}_2$ index. In this interpretation, the Dirac crossings on the $k_{x}=0,\pi$ lines correspond to gap closing transitions of the 1D Hamiltonian at which the $\mathbb{Z}_2$ changes. In a similar way, Fig.~\ref{fig:buckled_zeeman}(b) shows the effect of a Zeeman field $B_y$.

\begin{figure}
	\includegraphics[width=\columnwidth]{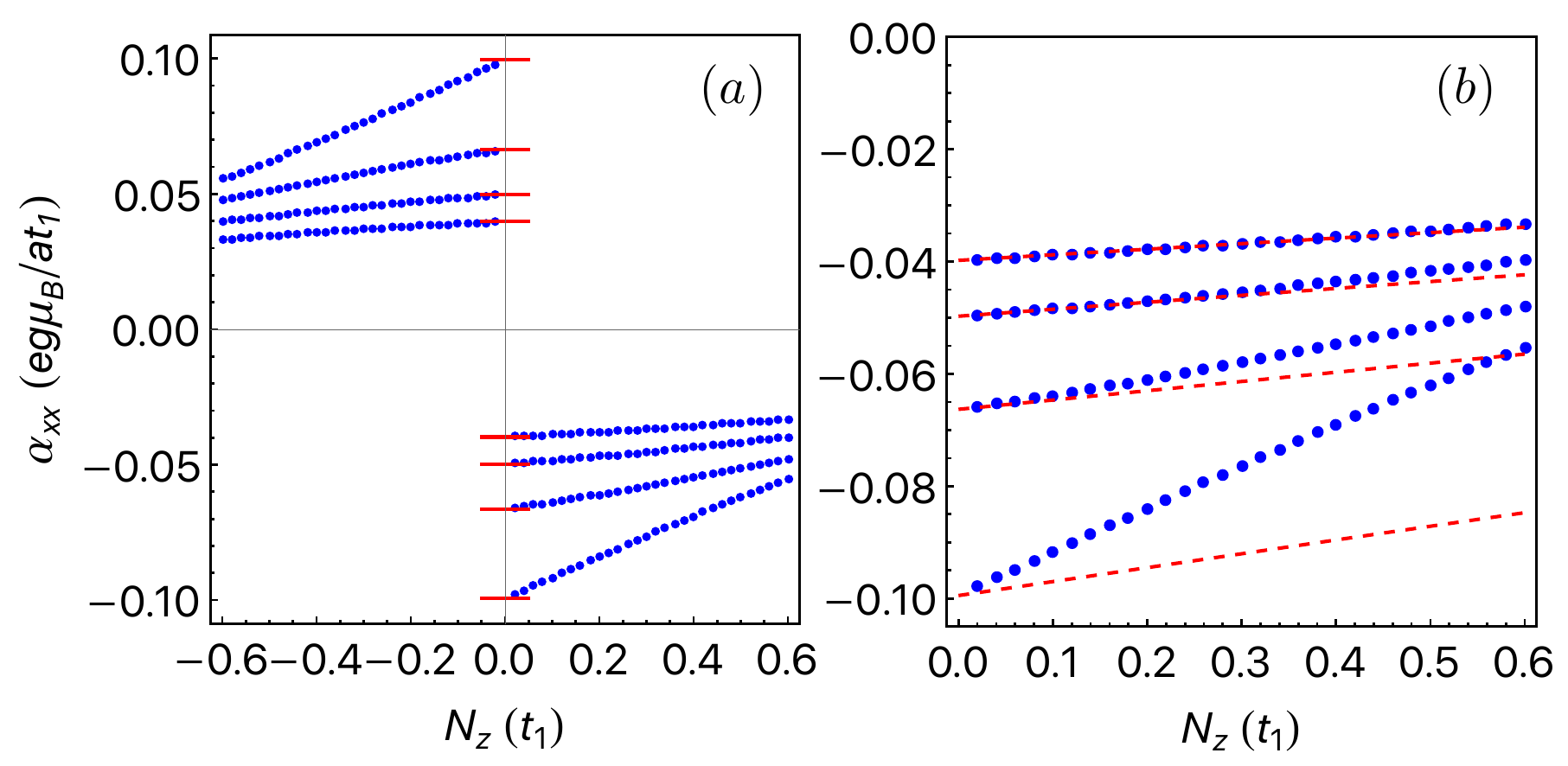}
	\caption{(a) Magnetoelectric polarizability $\alpha_{xx}=\alpha_{yy}$ of the buckled square model, calculated using Eq.~\eqref{eq:SMP-4band} and shown as a function of N\'eel order parameter $N_z$ (in units of $t_1$). Different curves correspond to $\lambda/t_1=0.4,0.6,0.8,1.0$, where larger values of $\lambda$ correspond to smaller (absolute) values of $\alpha_{xx}$. (b) Same as in (a) but only for $N_z/t_1>0$. The red dashed curves indicate analytical approximations obtained from considering only contributions from the continuum Dirac models at the high symmetry points.}
	\label{fig:alpha_xx_buckled}
\end{figure}

\subsection{Magnetoelectric polarizability of N\'eel antiferromagnet and multipolization\label{ssec:SMP-buckled}}

We then proceed to examining the SMP of the buckled square lattice antiferromagnet based on Eq.~\eqref{eq:SMP-4band}. It follows from the symmetry analysis of Sec.~\ref{ssec:square-landau} that when $\bN = N_z \hat \bz$, the nonzero components of $\alpha_{ij}$ are $\alpha_{xx}=\alpha_{yy}$. In Fig.~\ref{fig:alpha_xx_buckled}(a) we show $\alpha_{xx}$, numerically evaluated using Eq.~\eqref{eq:SMP-4band}, as a function of N\'eel order parameter $N_z$, and for different values of the spin-orbit coupling strength $\lambda$. The key feature evidenced by Fig.~\ref{fig:alpha_xx_buckled}(a) is the discontinuity of $\alpha_{xx}$ at $N_z=0$. This behavior is at odds with the naive expectation that $\alpha_{xx}$ should tend to zero linearly as $N_z\rightarrow 0$, since inversion and time-reversal symmetry are restored at $N_z=0$. The Landau theory of Sec.~\ref{ssec:square-landau}, which rests on symmetry arguments, supports this expectation~\footnote{Note that precisely at $N_z=0$ a microscopic calculation of the spin magnetoelectric polarizability is undefined since the bands cross and the system is not an insulator. Therefore, $\alpha_{xx}$ is calculated for finite but arbitrarily small $N_z$ (i.e., arbitrarily small Dirac masses), such the the limit $N_z\rightarrow 0$ can be examined and compared to the expectation from Landau theory.
}. 

The discontinuity at $N_z=0$ is caused by the Dirac points located at the high-symmetry points on the Brillouin zone boundary. Based on the analysis of Sec.~\ref{sec:general}, each of the Dirac points gives rise to a contribution $\propto \text{sgn}(N_z)$ to $\alpha_{xx}$, which is indeed what the calculation of $\alpha_{xx}$ shown in Fig.~\ref{fig:alpha_xx_buckled}(a) confirms. Hence, we find that the buckled square lattice antiferromagnet provides a concrete microscopic realization of the quasitopological magnetoelectric effect originating from 2D Dirac points. Here it is important to emphasize that this relies on the protection of the fourfold Dirac points by nonsymmorphic space group symmetries in the nonmagnetic state. 

The precise value of the discontinuity observed in Fig.~\ref{fig:alpha_xx_buckled}(a), i.e., the value of $\alpha_{xx}$ as $N_z$ approaches $0$, can be determined from the continuum Dirac theory developed in Sec.~\ref{ssec:buckled-dirac}. For each of the three high symmetry points $M$, $X$, and $X'$, we determine the shift $b_y$ in response to $B_x$, which, when summed, yields
\be
b_y = b^M_y + b^X_y+b^{X'}_y = -\frac{\bar{B}_x}{\hbar v^M}+\frac{\bar{B}_x}{\hbar v^X_x}-\frac{\bar{B}_x}{\hbar v^{X'}_y}. \label{eq:b_y-buckled}
\ee
Here we have abbreviated $\bar{B}_x \equiv g\mu_B B_x/2$. Noting that $\hbar v^X_x = \hbar v^{X'}_y = 2(t_1^2+\lambda^2)^{1/2}$ we conclude that the contributions from $X$ and $X'$ cancel, such that the topological 
contribution to $\alpha_{xx}$ is entirely due to $M$. According to the analysis of Sec.~\ref{ssec:em-dirac}, the SMP $\alpha_{xx}$ is given by
\be
 \frac{\partial P_x}{\partial B_x} = \text{sgn}(N_z) \frac{e}{2\pi }\frac{\partial b_y}{\partial B_x}= -\text{sgn}(N_z) \frac{e g\mu_B}{8\pi a \lambda },
\ee
where we have reinstated the lattice constant $a$. This value is indicated by red markers in Fig.~\ref{fig:alpha_xx_buckled}(a), showing full agreement with the numerical result determined using Eq.~\eqref{eq:SMP-4band}.

\begin{figure}
	\includegraphics[width=\columnwidth]{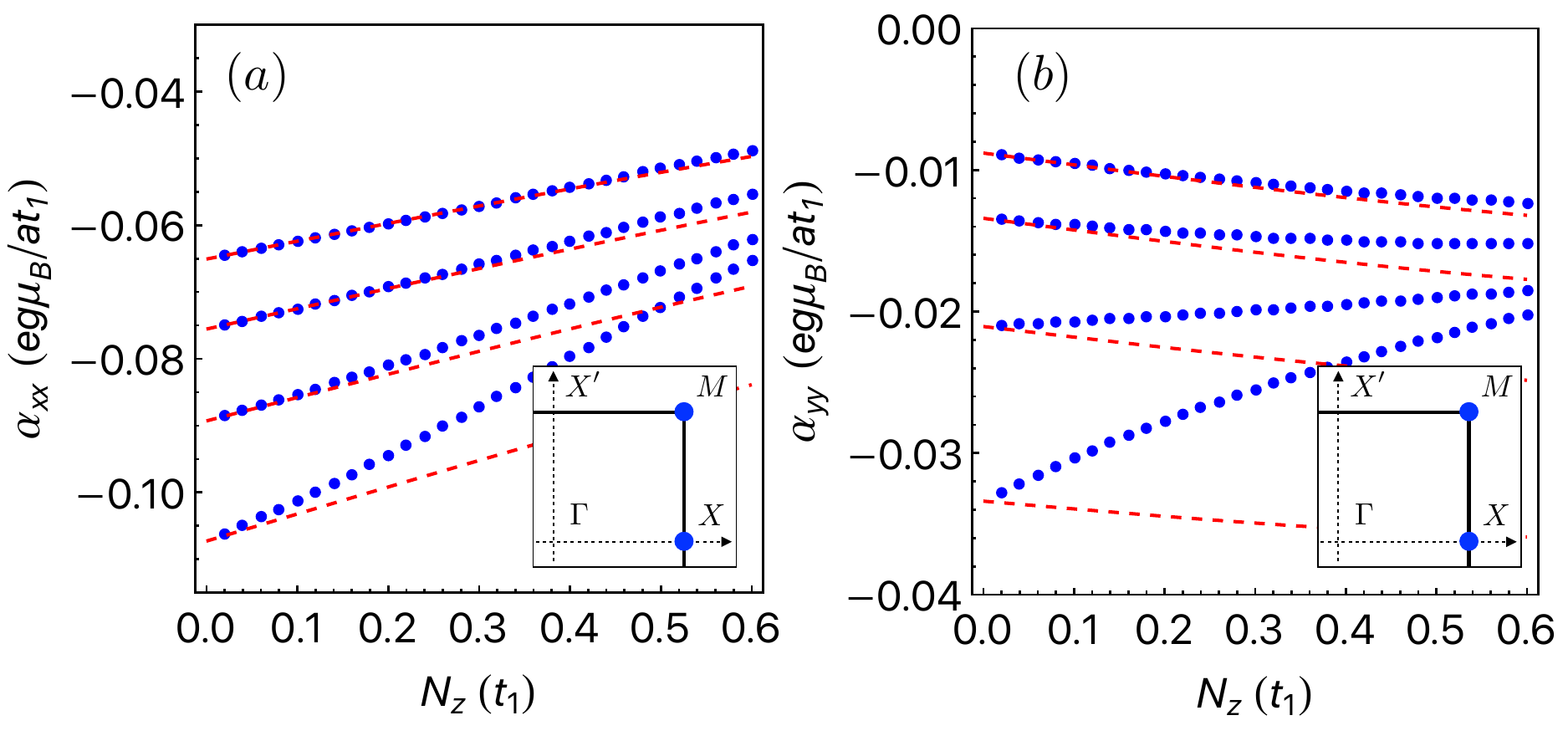}
	\caption{Magnetoelectric polarizabilities (a) $\alpha_{xx}$ and (b) $\alpha_{yy}$ of the buckled square lattice model when the symmetry breaking perturbation of Eq.~\eqref{eq:delta-H_k-buckled} is included. The symmetry breaking term gaps the Dirac point at $X'$, but preserves the Dirac points at $X$ and $M$, is indicated in the insets. In both panels the different curves correspond to $\lambda/t_1=0.4,0.6,0.8,1.0$, and we have set $\delta t =0.4t_1$. As in Fig.~\ref{fig:alpha_xx_buckled}(b), the blue dotted curves are obtained numerically using Eq.~\eqref{eq:SMP-4band}, and the red dashed lines indicate analytical approximations obtained from a continuum Dirac model for the high symmetry points. }
	\label{fig:alpha_xx_yy_buckled}
\end{figure}

Apart from the discontinuity at $N_z=0$, which is of topological origin, $\alpha_{xx}$ exhibits a second significant feature: a linear dependence on $N_z$. This is more clearly revealed in panel (b) of Fig.~\ref{fig:alpha_xx_buckled}, in which only the $N_z>0$ behavior is shown. This linear dependence constitutes a non-topological background contribution to the polarizability and can be understood by considering a continuum model for the electronic states close to the zone center $\Gamma$. Expanding Eq.~\eqref{eq:H_k-square} around $\bk=0$ yields $\mathcal H^\Gamma_\bq = m \tau^x - \hbar v \tau^z(q_x\sigma^y-q_y\sigma^x)$, where $m=-4t_1$, which shows that $\Gamma$ can be viewed as a gapped Dirac point with velocity $\hbar v = 2\lambda$ and Dirac mass $m$. Computing $\alpha_{xx}$ for a massive Dirac point in the regime $|N_z/m| \ll 1$ yields $\alpha_{xx}\simeq eg\mu_BN_z/4\pi \hbar v |m|$. This linear approximation is shown in Fig.~\ref{fig:alpha_xx_buckled}(b) by the red dashed curves, and is seen to agree well with the full result for larger values of the spin-orbit coupling strength $\lambda$. The latter is naturally expected, since the interpretation of the zone center $\bk=0$ as a gapped Dirac point becomes meaningful when the energy scale of spin-orbit coupling $\lambda$ is comparable to, or larger than, $t_1$. 

We conclude from Fig.~\ref{fig:alpha_xx_buckled} that the microscopic calculation of $\alpha_{xx}$ for the buckled square lattice antiferromagnet is well-explained by a continuum Dirac model for the electronic bands close to the four high symmetry points. In particular, the buckled square lattice antiferromagnet exhibits the quasitopological magnetoelectric effect associated with Dirac fermions in 2D.

To further cement this conclusion, we add a symmetry-breaking perturbation to the nonmagnetic Hamiltonian of Eq.~\eqref{eq:H_k-square}, which is of the form~\cite{Young:2015p126803}
\be
\delta  H_\bk = 4\, \delta t \,c_{x/2}s_{y/2}  \tau^y,  \label{eq:delta-H_k-buckled}
\ee
and represents an orthorhombic distortion of the crystal lattice. Its effect is to gap the Dirac point at $X'=(0,\pi)$~\cite{Young:2015p126803}. [Recall our definitions $c_{i/2} \equiv \cos( k_i/2)$ and $s_{i/2} \equiv \sin( k_i/2)$.] The spectrum of the distorted nonmagnetic lattice thus has symmetry-enforced Dirac points only at $X$ and $M$ (see insets Fig.~\ref{fig:alpha_xx_yy_buckled}). In Fig.~\ref{fig:alpha_xx_yy_buckled}(a) and (b) we show $\alpha_{xx}$ and $\alpha_{yy}$, respectively, of the symmetry-reduced system as a function of $N_z$ (only $N_z>0$ is shown).  Both polarizabilities, which are no longer equal as a result of the reduced symmetry, show the discontinuity at $N_z$. The topological contribution to the polarizabilities giving rise to this discontinuity now originates from two Dirac points, one located at $X$ and one at $M$, which follows from an analysis similar to Eq.~\eqref{eq:b_y-buckled}. The dashed red curves again show the continuum model approximations obtained from expansions around the four high symmetry points $\Gamma$, $X$, $X'$, and $M$, two of which are now described by gapped Dirac fermions. As in Fig.~\ref{fig:alpha_xx_buckled}(b), excellent agreement with the full numerical result is observed for larger values of $\lambda$.

\begin{figure}
	\includegraphics[width=\columnwidth]{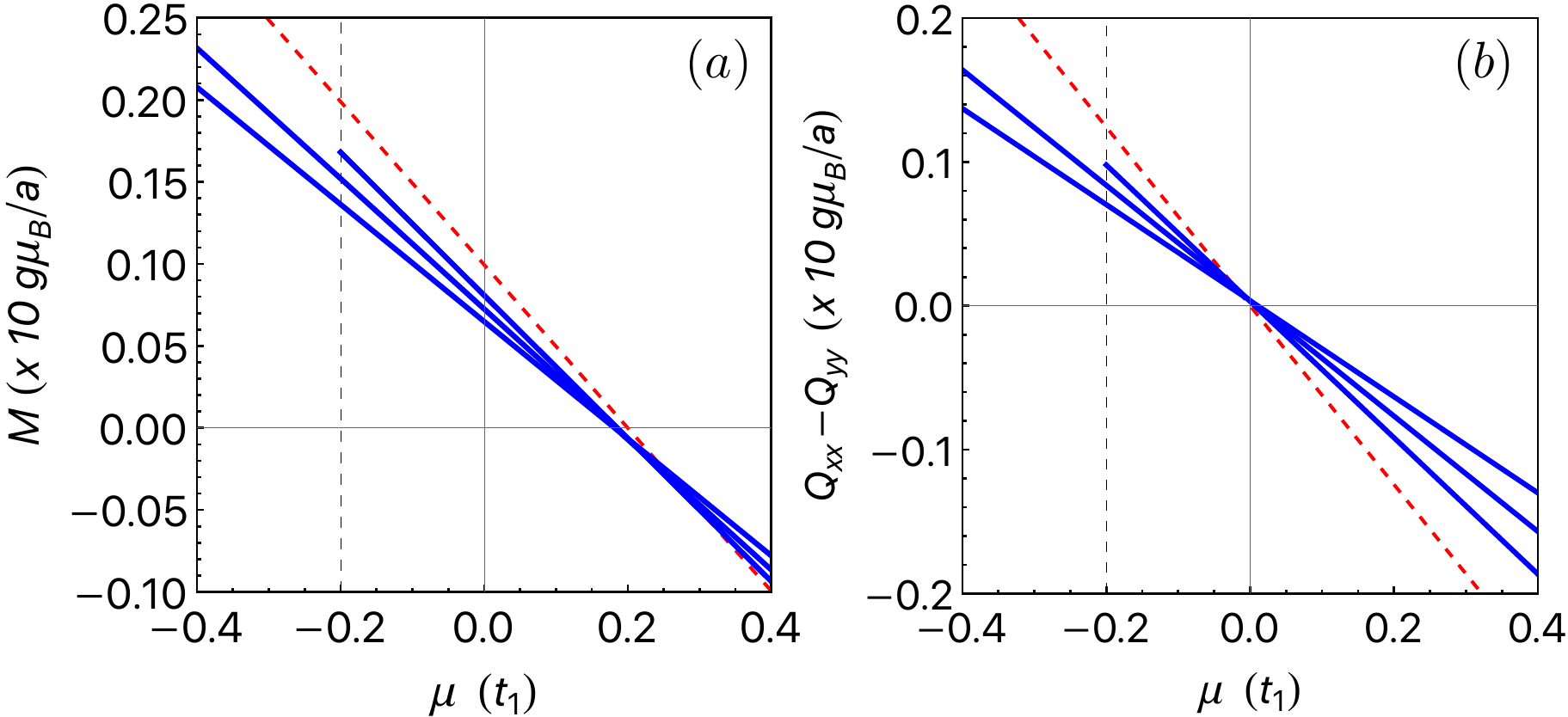}
	\caption{(a) Spin monopolization $M = (O_{xx}+O_{yy})/2$ of the tetragonal buckled square lattice model [Eqs.~\eqref{eq:H_k-square} and \eqref{eq:H_N,B}], calculated using Eq.~\eqref{eq:Oij-4band} and shown as a function of chemical potential $\mu$. Here we have set $\lambda = 0.8t_1$ and $t_2=0.05t_1$; blue lines correspond to $N_z/t_1=-0.4,-0.6,-0.8$ (top to bottom). The dashed red curve shows the result obtained from Dirac model, as given by Eq.~\eqref{eq:O_ij-Dirac2D}. The vertical dashed line marks the edge of the valence band when $N_z =0.4$. (b) Spin quadrupolization $O_{xx}-O_{yy}$ of the distorted buckled square lattice model, i.e., with nonzero $\delta t$ introduced in Eq.~\eqref{eq:delta-H_k-buckled}. We have set $\lambda = 0.8t_1$, $t_2=0.05t_1$, and $\delta t = 0.4t_1$. As in (a), the blue lines correspond to $N_z/t_1=-0.4,-0.6,-0.8$ (top to bottom) and the dashed red curve shows the result obtained applying Eq.~\eqref{eq:O_ij-Dirac2D} to $M=(\pi,\pi)$ and $X=(\pi,0)$. }
	\label{fig:multipolization}
\end{figure}

Up to this point, we have focused our microscopic analysis on the magnetoelectric polarizability $\alpha_{ij}$. In the final part of this section we briefly examine the closely related spin multipolization, as given by Eq.~\eqref{eq:Oij-4band}. In Fig.~\ref{fig:multipolization}(a), we show the spin monopolization, here defined as $M = (O_{xx}+O_{yy})/2$, for the tetragonal square lattice model (i.e., without distortions) as a function of the chemical potential $\mu$. To compute $O_{ij}$ using Eq.~\eqref{eq:Oij-4band}, we have introduced a uniform hopping term $\varepsilon_{0,\bk} =  -2 t_2(c_x+c_y)$, where $t_2$ is a second-nearest neighbor hopping. The blue curves correspond to three different values of $N_z$, specifically $N_z/t_1=0.4,0.6,0.8$, and the red dashed curve corresponds to the result obtained from applying Eq.~\eqref{eq:O_ij-Dirac2D} to the Dirac point at $(\pi,\pi)$. As expected, the calculated monopolization depends linearly on $\mu$ and approaches the result from the Dirac model as $N_z$ decreases. Note that the edge of the valence band is located at energy $-N_z +4t_2$, and since we are only interested in the insulating regime, we do not compute $M$ for values of $\mu$ below the valence band edge. 

In Fig.~\ref{fig:multipolization}(b) we show the spin quadrupolization $O_{xx}-O_{yy}$ obtained when the orthorhombic distortion given by Eq.~\eqref{eq:delta-H_k-buckled} is included. The general behavior of the quadrupolization is similar to monopolization, and in particular also approaches to result from the Dirac model (the red dashed line) as $N_z$ decreases. Note, however, that the Dirac model approximation of the quadrupolization is proportional to $\mu$ and vanishes for $\mu=0$, since it is only determined by the Dirac point at $X$, which sits at energy zero [$\varepsilon_D=0$ in Eq.~\eqref{eq:O_ij-Dirac2D}].

\section{Square lattice double-$Q$ spin model \label{sec:noncollinear}}

The purpose of this section is to demonstrate the topological spin magnetoelectric effect in a second microscopic lattice model for a 2D antiferromagnet. This second model is also defined on a square lattice, but differs in two important ways from the buckled square lattice considered in the previous section. First, spin-orbit coupling is neglected, such that the model has a full spin-rotation symmetry, and second, the magnetic state is non-collinear and defined by a coplanar double-$Q$ configuration of ordered moments (see Fig.~\ref{fig:figvdbmodel}). 
The remainder of this section will resemble the discussion of the buckled lattice model. We first describe the symmetries and energy bands of our spin-orbit-free model, then examine the allowed magnetoelectric couplings, and finally compute the magnetoelectric polarizability highlighting the occurrence of the topological spin multipolization.

\subsection{Model and symmetries}

\begin{figure}[tbp]
    \begin{center}
         \includegraphics[width=\columnwidth]{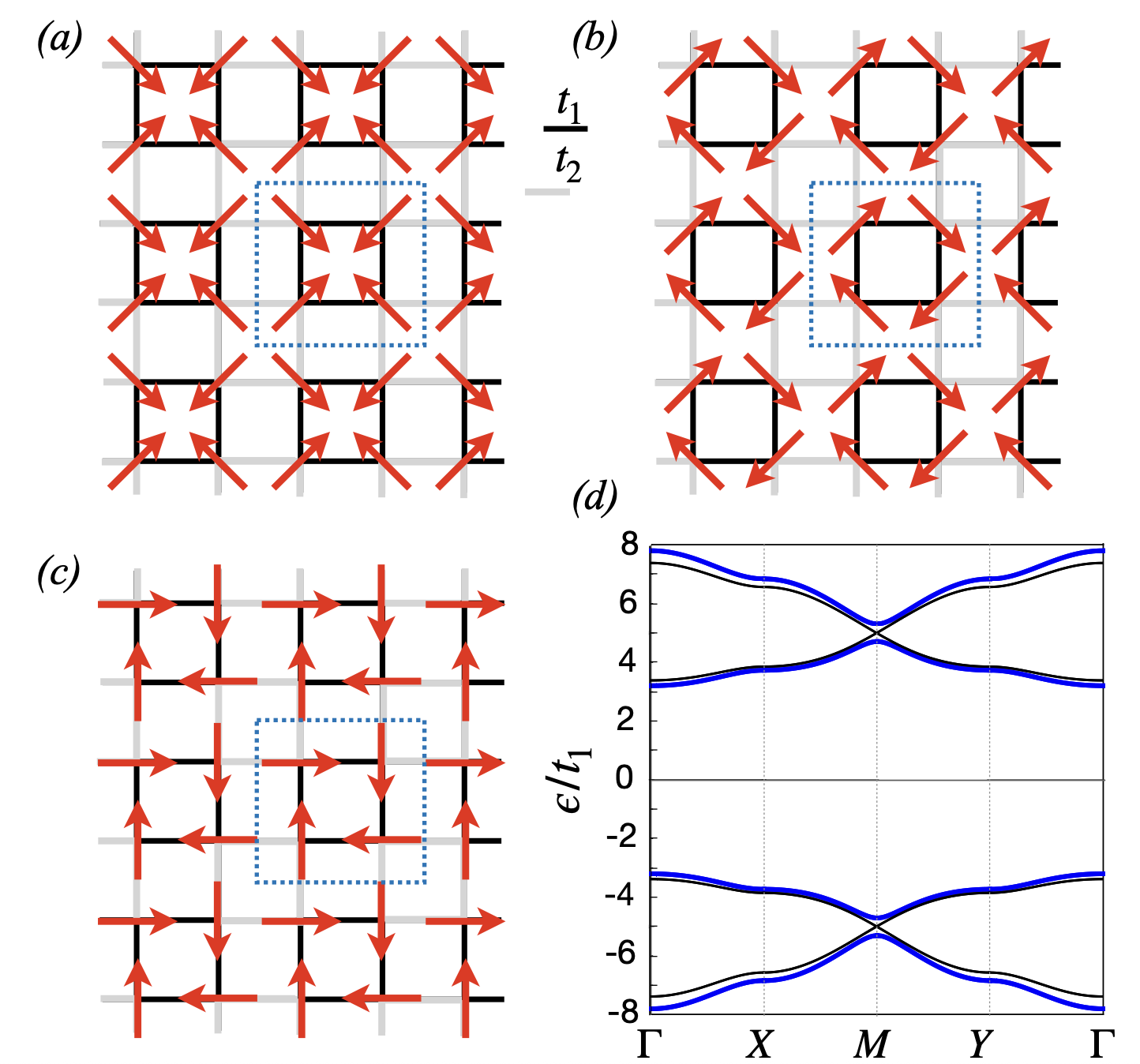}
    \caption{(a),(b),(c) Sketch of the square lattice double-$Q$ spin model. $t_1$ and $t_2$ are the two alternating nearest neighbor bond strengths. The dashed blue box indicate the lattice unit cell. The arrows on each lattice sites indicate the directions of the classical spin. The three configurations in (a),(b),(c) are characterized by different but related spin multipoles. (d) Electronic band structure for a Kondo coupling $J=5~t_1$. For $t_2=t_1$ (black thin lines) the model realizes a Dirac antiferromagnet. For $t_2=1.3 t_1$ (blue thick lines) the Dirac cones are gapped out by the bond density waves.}
    \label{fig:figvdbmodel}
    \end{center}
\end{figure}

The model considered in this section describes itinerant electrons on a simple square lattice coupled to an ordered configuration of local moments. The tight-binding Hamiltonian of this model reads as
\begin{equation}
{\mathcal H}=\sum_{\langle i j \rangle} t_{ \langle i j \rangle} c^{\dagger}_i c_j  + J  \sum_i {\bf S}_i \cdot  c^{\dagger}_i \boldsymbol{\sigma} c_i, 
\label{eq:modelvanderbilt}
\end{equation}
where ${\bf S}_i$ are classical local moment spins (of unit length), $J$ is a Kondo coupling between electrons and spins, and $t_{ \langle i j \rangle} $ are nearest neighbor hopping amplitudes. The ordered configuration of local moments is given by a double-$Q$ spin state generally written as ${\bf S}_i={\bf S}_1 \mathrm{e}^{i {\bf Q}_1 \cdot {\bf r}_i} + {\bf S}_2 \mathrm{e}^{i {\bf Q}_2 \cdot {\bf r}_i} $, with wave vectors ${\bf Q}_1=\left(\pi,0 \right)$ and ${\bf Q}_2=\left(0, \pi \right)$. Here ${\bf S}_1 $ and ${\bf S}_2$ are the corresponding Fourier components, which must be real and on which we impose 
the fixed-length constraint ${\bf S}_1 \cdot {\bf S}_2=0$. Three different realizations of the double-$Q$ order are shown in panels (a--c) of Fig.~\ref{fig:figvdbmodel}. The configuration of panel (a) corresponds to the choice ${\bf S}_1 = \hat \bx$ and ${\bf S}_2=\hat \by$, whereas panel (b) corresponds to the choice ${\bf S}_1 = \hat \by$ and ${\bf S}_2=-\hat \bx$. Finally, panel (c) corresponds to ${\bf S}_1=(\hat \bx + \hat \by)/\sqrt{2}$ and ${\bf S}_2=(\hat \bx - \hat \by)/\sqrt{2}$. All three configurations are related by a global spin rotation and are therefore energetically equivalent in the absence of spin-orbit coupling.

We further assume a staggered modulation of the nearest neighbor hoppings in both the $\hat x$ and $\hat y$ directions, as shown in Fig.~\ref{fig:figvdbmodel}. The two alternating hopping amplitudes are labeled $t_1$ and $t_2$. The electronic energy bands obtained from Eq.~\eqref{eq:modelvanderbilt} are shown in Fig.~\ref{fig:figvdbmodel}(d), both without hopping modulations ($t_1=t_2$; thin black curves) and with hopping modulations (thick blue curves). The spectrum consists of two branches separated by an energy $J$, which we assume is the largest energy scale. We focus our discussion on the lower branch. First note that all bands are twofold degenerate due to $\mathcal I \mathcal T$ symmetry, with the inversion center located at the center of the square plaquette. Without hopping modulations the bands cross at $M=(\pi,\pi)$, realizing a fourfold degenerate Dirac point protected by the symmetries of double-$Q$ spin state~\cite{Affleck:1988p3774,Agterberg:2000p13816}. Specifically, in the absence of hopping modulations the spin state shown in Fig.~\ref{fig:figvdbmodel}(a) has two generalized fractional translation symmetries given by $\mathcal U_{2y}  \{ \mathbb{1}  | \tfrac{1}{2} 0\} $ and $\mathcal U_{2x}  \{ \mathbb{1}  | 0 \tfrac{1}{2} \} $, where $\mathcal U_{2y}$ and $\mathcal U_{2x}$ are pure spin rotations (i.e., rotations of the electron spin only) and the translations are defined with respect to basis vectors of the magnetic unit cell. The spin state is furthermore left invariant by the product of time-reversal $\mathcal T$ and the diagonal fractional translation $\{ \mathbb{1} | \tfrac{1}{2} \tfrac{1}{2}\} $. The algebraic properties of these symmetries have the following implications at $M=(\pi,\pi)$: the two fractional translations mutually anticommute and square to $+1$, whereas the antiunitary symmetry $\mathcal T \{ \mathbb{1} | \tfrac{1}{2} \tfrac{1}{2}\} $ squares to $-1$ and commutes with the translations. This implies that all bands must be fourfold degenerate at $M$ (see Appendix~\ref{app:P4/nmm} for details on space group algebra and degeneracies). 

The Dirac point at $M$ acquires a gap in the presence of bond modulations, which break the generalized translation symmetries~\cite{Seradjeh:2008p033104}. The double-$Q$ spin state with bond modulations thus realizes a gapped Dirac theory similar to the antiferromagnetic buckled square lattice model of the previous section, with the notable difference that spin-orbit coupling is absent and that the bond modulations provide the Dirac mass. Below, we will focus on the insulating state which arises at quarter filling. All results equally apply to the insulating state at three-quarter filling.

The structure of the allowed magnetoelectric response of the insulating antiferromagnet can be inferred from an analysis of its crystalline symmetries. Below we determine the magnetoelectric response using a Landau theory approach similar to Sec.~\ref{ssec:square-landau}, but as a first step towards such a more general approach it is instructive to analyze the specific spin configuration of Fig.~\ref{fig:figvdbmodel}(a). Choosing the origin at the center of the unit cell (i.e., at the center of the dashed square; see Fig.~\ref{fig:figvdbmodel}), it is clear that the product ${\mathcal U}_{2x} {\mathcal C}_{x}$ leaves the state invariant, where ${\mathcal C}_{x}$ denotes a spatial twofold rotation and ${\mathcal U}_{2x}$ is a pure spin rotation. (Recall that spin-orbit coupling is absent.) The magnetic state is also left invariant by ${\mathcal U}_{2y} {\mathcal C}_{y}$ and ${\mathcal U}_{2z} {\mathcal C}_{z}$, and the combination of these three symmetries forces all off-diagonal components of $\alpha_{ij}$ to vanish. Off the diagonal components, $\alpha_{zz}$ must vanish due to mirror reflection symmetry in the plane, and $\alpha_{xx}$ must equal $ \alpha_{yy}$ due to fourfold rotation symmetry. This symmetry analysis thus show that the magnetoelectric response is governed by one independent parameter.

Although the symmetry analysis is specific to the spin configuration shown in Fig.~\ref{fig:figvdbmodel}(a), the spin configurations shown in panels (b) and (c) are related to (a) by a global spin rotation, and the corresponding magnetoelectric responses thus follow directly. This is most clearly exposed by the Landau theory approach, which we now turn to.

\subsection{Landau theory of magnetoelectric response \label{ssec:double-Q-landau}}

As demonstrated in Sec.~\ref{ssec:square-landau}, Landau theory provides a way to frame magnetoelectric responses in very general terms. Whereas in Sec.~\ref{ssec:square-landau} we determined all symmetry-allowed trilinear couplings between polarization, magnetization, and N\'eel order, in the present case an appropriate Landau theory must account for two distinct orders. In addition to the magnetic order parameters ${\bf S}_1 $ and ${\bf S}_2$ one must also consider the bond density wave modulations of the hopping amplitudes. The necessity of bond density wave order (i.e., the hopping modulations) follows from the observation that uniform hopping implies an inversion symmetry with the lattice sites as inversion centers. Within the framework of Landau theory we therefore consider a simple square lattice which develops two distinct types of order: spin order and bond density wave order. Our goal is to determine all symmetry-allowed quadrilinear coupling between these orders and the polarization and magnetization.

We introduce general bond density wave order parameters by writing the staggered hopping patterns as
\be
\delta t_{i,i+\hat x}=\Delta_1 \mathrm{e}^{i {\bf Q}_1 \cdot {\bf r}_i}, \quad \delta t_{i,i+\hat y}=\Delta_2 \mathrm{e}^{i {\bf Q}_2 \cdot {\bf r}_i}. \label{eq:bond-order}
\ee
Here ${\bf Q}_1$ and ${\bf Q}_2$ are the same wave vectors that define the spin order, and $\Delta_1$ and $\Delta_2$ are the corresponding order parameters. Note that in the model discussed above have made the choice $\Delta_1=\Delta_2 = \delta t /2$, such that $t_1 = \bar t + \delta t /2$ and $t_2 = \bar t - \delta t /2$, but in general these are independent order parameters. To construct the required quadrilinear couplings, we must require ({\it i}) translational invariance with respect to the underlying square lattice and  ({\it ii}) full rotation invariance in spin space (since spin-orbit coupling is absent). Imposing these two symmetries gives rise to the allowed couplings
\begin{multline}
{\mathcal F}_{\Delta,S,P,M}= \beta_{xx} \Delta_1 P_x   \left({\bf S}_1 \cdot {\bf M}  \right) + \beta_{xy} \Delta_1 P_y  \left({\bf S}_1 \cdot {\bf M}  \right)   \\
+ \beta_{yx} \Delta_2 P_x  \left({\bf S}_2 \cdot {\bf M}  \right) + \beta_{yy} \Delta_2 P_y   \left({\bf S}_2 \cdot {\bf M}  \right).
\end{multline}
Note that translational invariance requires that $\Delta_{1,2} $ and ${\bf S}_{1,2}$ are coupled, since both carry wave vector ${\bf Q}_{1,2}$. Further imposing the rotation symmetries of the square lattice reduces the allowed couplings to a single term given by
\begin{equation}
{\mathcal F}_{\Delta,S,P,M}=\beta \left[\Delta_1 P_x \left({\bf S}_1 \cdot {\bf M}  \right)+ \Delta_2 P_y \left({\bf S}_2 \cdot {\bf M} \right)\right] . \label{eq:landau-double-Q}
\end{equation}
From this Landau theory the magnetoelectric response can be determined for a given spin configuration and a given realization of hopping modulations. Global spin rotation invariance is manifest in Eq.~\eqref{eq:landau-double-Q} and it is clear how the magnetoelectric responses of two spin states that differ by a global spin rotation are related.

Let us then apply the result of \eqref{eq:landau-double-Q} to the specific order parameter configurations shown in Fig.~\ref{fig:figvdbmodel}. As mentioned, in Fig.~\ref{fig:figvdbmodel}(a) we have chosen ${\bf S}_1 = \hat \bx$ and ${\bf S}_2=\hat \by$, and furthermore $\Delta_1=\Delta_2 = t_1-t_2=\delta t$. Inserting this in \eqref{eq:landau-double-Q} yields
\begin{equation}
{\mathcal F}_{P,M}=\beta \delta t \left(P_x M_x+  P_y M_y \right),
\label{eq:landau-double-Q-monopole}
\end{equation}
and from this result it follows that the only allowed components of $\alpha_{ij}$ are $\alpha_{xx}=\alpha_{yy}$. Consequently, the magnetic configuration of Fig.~\ref{fig:figvdbmodel}(a) is associated with a magnetoelectric monopole. Instead, the spin configuration of Fig.~\ref{fig:figvdbmodel}(b) corresponds to ${\bf S}_1 = \hat \by$ and ${\bf S}_2=-\hat \bx$. Inserting this in \eqref{eq:landau-double-Q} yields
\begin{equation}
{\mathcal F}_{P,M}=\beta \delta t \left(P_x M_y-  P_y M_x \right),
\label{eq:landau-double-Q-toroidal}
\end{equation}
which reveals that in this case $\alpha_{ij}$ is fully off-diagonal and antisymmetric. The magnetic configuration of Fig.~\ref{fig:figvdbmodel}(b) is thus characterized by a toroidal moment. The spin configuration of Fig.~\ref{fig:figvdbmodel}(c) is intermediate between these two cases and gives rise to both diagonal and off-diagonal components.

\begin{figure}[tbp]
    \begin{center}
         \includegraphics[width=\columnwidth]{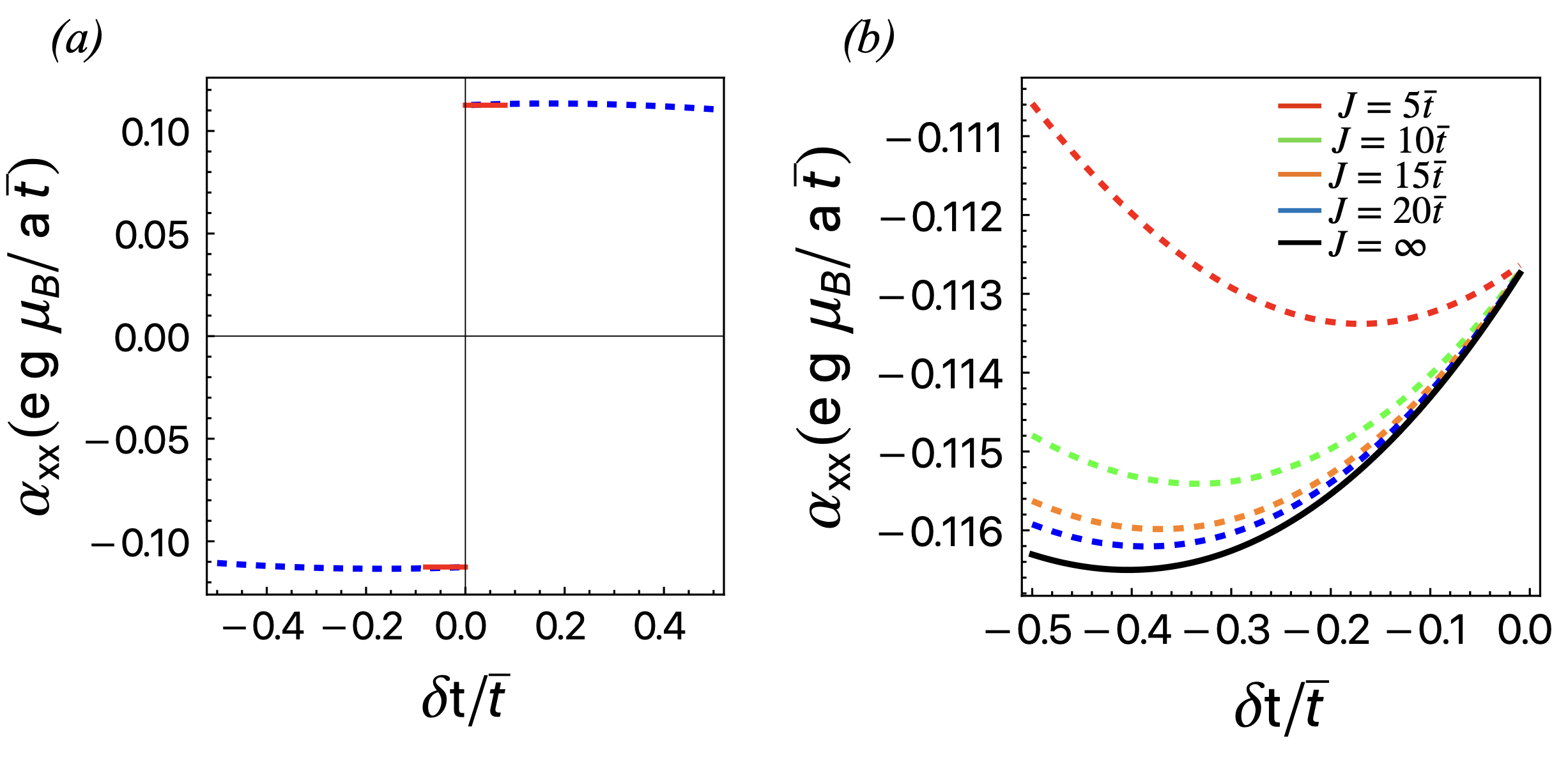}
    \caption{(a) Magnetoelectric polarizability $\alpha_{xx}$ of the square lattice double-$Q$ spin model at one-quarter filling as a function of the bond density wave order parameter $\delta t=t_1-t_2$ measured in units of $\bar{t}=(t_1+t_2)/2$. The Kondo coupling constant $J=5 \bar{t}$. The red bar indicates the analytical value obtained using a Dirac theory. (b) Close-up of (a) showing the $\delta t$ dependence at different values of the Kondo coupling constant. For $J \gg \bar{t}$ the behavior is well captured by the ($J=\infty$) effective four-band tight-binding model of Eq.~\ref{eq:loweneTBvdbmodel}. 
    }
    \label{fig:figvdbSMP}
    \end{center}
\end{figure}

\subsection{Magnetoelectric polarizability and continuum Dirac theory}

We next proceed by determining the magnetoelectric microscopically, specifically for the magnetic order parameter configuration of Fig.~\ref{fig:figvdbmodel}(a). A numerical 
calculation based on Eq.~\eqref{eq:SMP} for the full eight-band model confirms the existence of two non-vanishing and equal SMP components $\alpha_{xx}=\alpha_{yy}$. In Fig.~\ref{fig:figvdbSMP}(a) we show our numerical results for $\alpha_{xx}$ as a function of $\delta t$. 
Precisely as for the buckled square lattice model considered in Sec.~\ref{sec:square}, the key feature evidenced in Fig.~\ref{fig:figvdbSMP}(a) is the discontinuity of $\alpha_{xx}$ at $\delta t=0$. This is in contradiction with the Landau theory analysis discussed above, in particular in Eq.~\ref{eq:landau-double-Q-monopole}, which would instead suggest that $\alpha_{xx}$ tends to zero as $\delta t \rightarrow 0$. The discontinuity of the SMP component $\alpha_{xx}$ at $\delta t=0$ is again caused by the existence of Dirac crossings, in particular the Dirac point located at the $M$ point mentioned above. This can be shown by determining the effective long-wavelength continuum theory of the model given by Eq.~\eqref{eq:modelvanderbilt}.

To do so, we first consider the system at $t_1 \equiv t_2$, in which case the Dirac point is gapless, and choose a basis in which the Kondo coupling term with coupling $J$ is diagonal. We focus on the sector with energy $-J$ and project out the $+J$ sector, which yields an effectively spinless four-band model (since spin is tied to the local moments). Expanding in the small momentum $\bq$ and performing two additional rotations in the low-energy sector, we obtain a Dirac theory that can be written in terms of four Dirac matrices as
\begin{equation} 
{\mathcal H}_{\bf q}^M=\hbar v_F \left(-q_x \tilde\sigma_z \tilde\tau_z + q_y \tilde\sigma_y \tilde\tau_0 \right) + \dfrac{\delta t}{\sqrt{2}} \left( -\tilde\sigma_x \tilde\tau_0 + \tilde\sigma_z \tilde\tau_y \right)
\label{eq:hamlowenergydoubleQ}
\end{equation}
where the Dirac mass $\delta t=t_1-t_2$, while the Fermi velocity $v_F= a~ {\bar t} /(\sqrt{2} \hbar)$ and ${\bar t}=\left(t_1+t_2\right)/2$. We have used a tilde to denote the Pauli matrices $\tilde{\sigma}_i$ and $\tilde{\tau}_i$, to emphasize that these do not directly correspond to spin or (magnetic) sublattice degrees of freedom, but instead are introduced to span the low-energy subspace. To compute the SMP it is necessary to determine the form of the physical spin operators when projected into the low-energy subspace. We find that in the chosen basis the spin operators $\tilde s_{x,y}$ can be expressed in terms of commutators of Dirac matrices 
\begin{eqnarray*}
\tilde s_x&=& -\dfrac{1}{2 \sqrt{2} i} \left[\tilde \sigma_y \tilde\tau_z , \tilde\sigma_z \tilde\tau_z \right] \\
\tilde s_y&=& -\dfrac{1}{2 \sqrt{2} i} \left[\tilde\sigma_x \tilde\tau_0 , \tilde\sigma_y \tilde\tau_0  \right]. 
\end{eqnarray*}
In the massless case, {\it i.e.} for $\delta t=0$, the effect of a Zeeman coupling due to an external magnetic field will be to shift the twofold Dirac cones in opposite directions along the BZ boundary. Specifically, the Dirac cones are shifted along $X-M$ or $Y-M$ line for a magnetic field along the ${\hat x}$ and ${\hat y}$ direction, respectively. As discussed in Appendix~\ref{app:dirac-shift-double-Q} the shift of the Dirac crossings in response to a Zeeman field is protected by symmetry, and thus valid beyond the low-energy description. The situation is thus similar to the buckled square lattice model: in the massless limit, the Zeeman field enters as a pseudo-gauge field, shifting the Dirac points in momentum space.

Using the expression for the SMP of a general four-band model, as derived and discussed in Appendix~\ref{app:four-band}, and applying it to the continuum Dirac model of Eq.~\eqref{eq:hamlowenergydoubleQ}, we obtain that the Dirac point gives rise to an SMP contribution $\alpha_{xx} =\alpha_{yy}= e g \mu_B \textrm{sgn}(\delta t) / (2 \sqrt{2} \pi a \bar{t})$. This value is indicated with the red markers in Fig.~\ref{fig:figvdbSMP}(a) and is in full agreement with the numerical result. The topological contribution to the SMP originating from the Dirac point at $M$ is the central result of this section and shows that the same fundamental physics identified in the buckled lattice model---a magnetic insulator described by a massive Dirac theory---is at play here.

Apart from the discontinuity due to the topological contribution to the SMP, the behavior of $\alpha_{xx}$ at finite $\delta t$ [see Fig.~\ref{fig:figvdbSMP}(b)] exhibits an additional weak dependence on the Dirac mass $\delta t$. It originates from two contributions of a different nature. The first contribution is from crystalline anisotropy terms in the low-energy manifold of four bands that warp the massive Dirac cone. The second stems from the hybridization between the two quartets of bands separated by the Kondo coupling $J$. 
We can effectively keep track of the former by projecting the Bloch Hamiltonian onto the low-energy manifold at energy $-J$ and retain its full momentum dependence (i.e., not expand in small momentum around $M$). After a momentum translation ${\bf k} \rightarrow {\bf k} + M$, such that $\bk =0$ corresponds to $M=(\pi,\pi)$, we obtain the effective four band Bloch Hamiltonian
\begin{eqnarray} 
{\mathcal H}_{LE}&=&\sqrt{2} \bar{t} \left(-\sin{\frac{k_x}{2}} \sigma_z \tau_z + \sin{\frac{k_y}{2}} \sigma_y \tau_0 \right) + \nonumber \\ & & \dfrac{\delta t}{\sqrt{2}} \left( -\cos{\frac{k_y}{2}}\sigma_x \tau_0 + \cos{\frac{k_x}{2}} \sigma_z \tau_y \right).
\label{eq:loweneTBvdbmodel}
\end{eqnarray}
This Hamiltonian can again be expressed in terms of four anticommuting Dirac matrices. An explicit computation using the method outlined in Appendix B then leads to the $J=\infty$ behavior reported in Fig.~\ref{fig:figvdbSMP}(b). The deviation from this behavior, which becomes more clearly visible by decreasing the Kondo coupling $J$, can be then attributed to the hybridization between the two quartets of bands.

\section{Toy models in one dimension \label{sec:1D}}

The discussion of the previous section highlights and emphasizes the observation that the origin of the (quasi)topological magnetoelectric polarizability of 2D Dirac semimetals can be traced back to the to the strong $\mathbb{Z}_2$ topological numbers characterizing 1D topological crystalline insulator. To further expose and examine the significance of this topological crystalline invariant in 1D, in this section we investigate two simplified 1D toy models derived from the microscopic 2D models considered in Secs.~\ref{sec:square} and \ref{sec:noncollinear}.

\subsection{Buckled 1D zigzag chain}

\begin{figure}
	\includegraphics[width=\columnwidth]{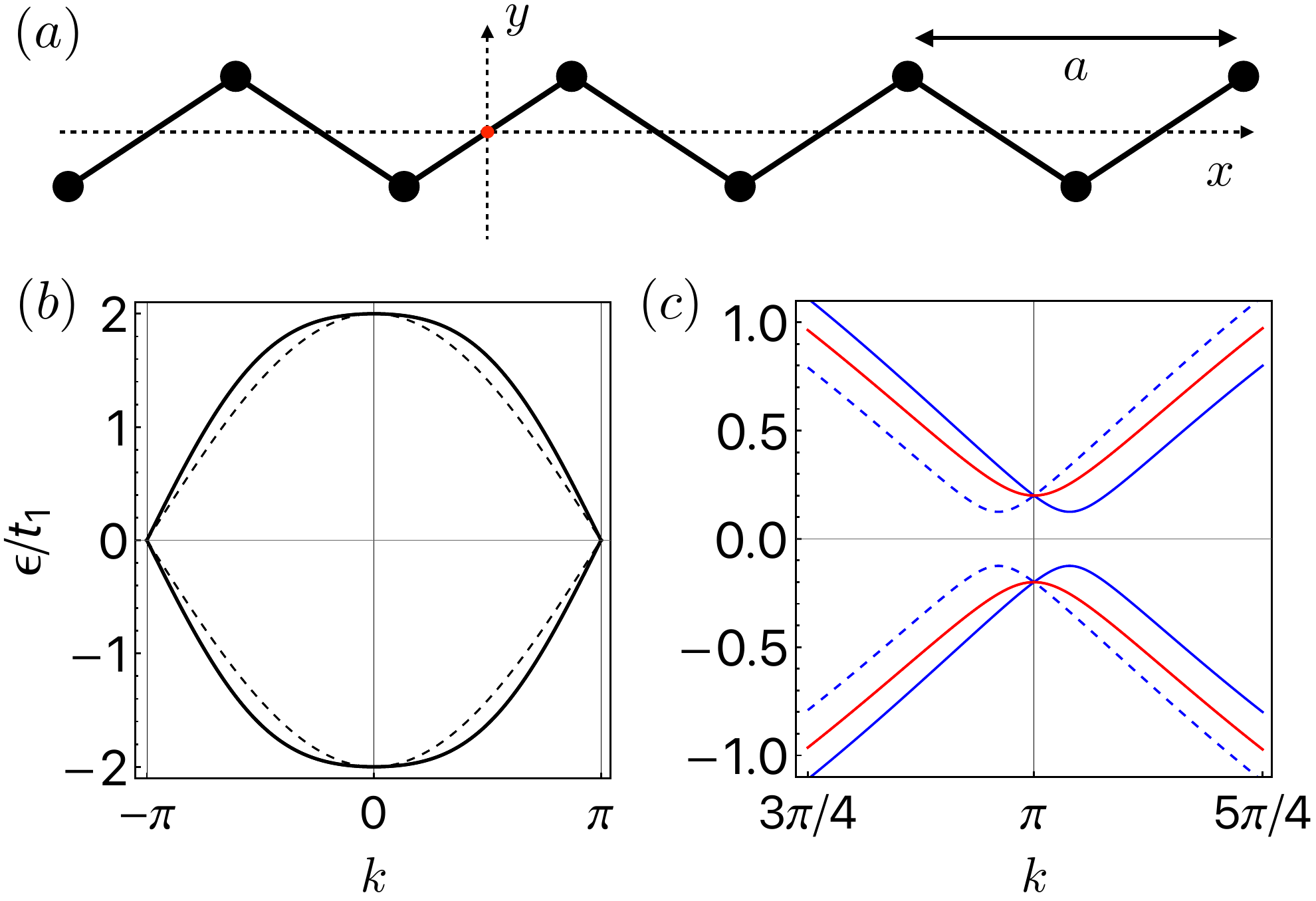}
	\caption{(a) Buckled ``zigzag'' chain in one dimension (1D) with inversion centers between the $A$ and $B$ sites; the lattice sites themselves are not inversion centers. (b) Energy spectrum of Eq.~\eqref{eq:H_k-1D-buckled} without N\'eel order and Zeeman field ($\bN = \bB =0$). Dashed and solid line correspond to $\lambda=0$ and $\lambda = 0.4t_1$, respectively. (c) Energy spectrum with $N_y = 0.2t_1$ (red curve) and $g\mu_B B_x/2 = 0.2$ (blue curves). In both cases $\lambda = 0.4t_1$. }
	\label{fig:zigzag}
\end{figure}

The first 1D toy model we consider is the buckled zigzag chain shown in Fig.~\ref{fig:zigzag}(a), which may be viewed as a 1D version of the 2D buckled square lattice. The buckling gives rise to two sublattices, again denoted $A$ and $B$, and using the framework of Eq.~\eqref{eq:H-def} a minimal Hamiltonian for the buckled chain can be expressed as~\cite{Yanase:2014p014703}
\be
H_{k} = -2 t_1 c_{k/2}  \tau^x  - \lambda s_k \tau^z\sigma^z +  \tau^z \bN \cdot \bsigma -\frac12 g \mu_B \bB\cdot \bsigma. \label{eq:H_k-1D-buckled}
\ee
Here the first term corresponds to nearest-neighbor hopping (between sublattices) and the second terms is a Kane-Mele-type spin-orbit coupling activated by the buckling~\cite{Kane:2005p226801}. The third and fourth terms represent N\'eel order and the Zeeman field, as before. Note that we have abbreviated $c_{k/2} \equiv \cos(k/2)$ and $s_k \equiv \sin k$, where $k$ is the momentum along the chain direction. The lattice constant is set to $a=1$. 

The point group $D_{2h}$ of the buckled chains is embedded in a nonsymmorphic space group generated by $ \{\mathcal C_{2x} | \tfrac12 \}$, $ \{\mathcal C_{2z} |0  \}$, and $ \{\mathcal I | 0 \}$. All three generators exchange the two sublattices, such that products of the generators do not exchange the sublattices. Examples of the latter are the twofold rotation $\{\mathcal C_{2y} | \tfrac12  \} $ or the mirror reflection $\{\mathcal M_x  | \tfrac12  \} $. The nonsymmorphic structure of the space group has important implications for the energy spectrum of $H_k$. The spectrum for $\bN = \bB =0$ (i.e. no N\'eel order or Zeeman field) is shown in Fig.~\ref{fig:zigzag}(b), where the dashed and solid curves correspond to $\lambda =0 $ and $\lambda = 0.4t_1$, respectively. Note that the energy bands are manifestly twofold degenerate due to the presence of both inversion and time-reversal symmetry. 

When the spin-orbit term is ignored, the energy bands simply correspond to the folded single-band spectrum of an ordinary 1D chain. The folding gives rise to a linear crossing at $k=\pi$, which must remain in the presence of SOC as a result of space group symmetry. In fact, the space group algebra mandates that all energy bands at $k=\pi$ must be fourfold degenerate. In this sense, the linearly dispersive fourfold degeneracy is equivalent to the symmetry-enforced Dirac points of the 2D buckled square lattice, as is detailed further in Appendix~\ref{app:1D-chain}. Expanding the Hamiltonian in small momentum $q$ in the vicinity of $k=\pi$ yields
\be
\mathcal H_q = t_1 q \tau^x + \lambda q \tau^z\sigma^z \rightarrow U^\dagger \mathcal H_q U = \hbar v q \tau^z\sigma^z,
\ee
where, as in Sec.~\ref{ssec:buckled-dirac}, we have performed a rotation $U=\exp(-i\theta \tau^y\sigma^z/2)$ to bring the expanded Hamiltonian into standard Dirac form, and have defined $v = (t^2_1+\lambda^2)^{1/2}/\hbar$ with $\tan\theta = t_1/\lambda$. 

Consider then the effect of antiferromagnetic N\'eel order and of a Zeeman field. For our purpose it is sufficient to analyze the case in which the N\'eel vector points in the $\hat y$ direction ($N_y$) and the Zeeman field points in the $\hat x$ direction ($B_x$). As is explained in more detail in Appendix~\ref{app:1D-chain}, N\'eel order in the $\hat y$ direction allows for a nonzero $\alpha_{xx}$, which we determine and examine below.

The effect of nonzero $N_y$ on the spectrum is shown in Fig.~\ref{fig:zigzag}(c) by the red curve. The bands remain twofold degenerate due to $\mathcal I \mathcal T$ symmetry, but the Dirac point at $k=\pi$ acquires a gap. In the resulting insulating state the electronic contribution to the polarization $P_x$, which may be calculated using the modern theory of polarization~\cite{Resta:1992p51,King-Smith:1992p1651,Vanderbilt:1993p4442,Resta:1994p899}, must vanish due to the presence of the combined $\mathcal I \mathcal T$ symmetry. This can also be understood by noting that any symmetry-adapted Wannier function~\cite{Soluyanov2011,Po2017,Bradlyn2017,vanmiert2018} will have a time-reversed partner centered at a Wyckoff position related by inversion. Therefore, the contribution to the electronic polarization of each time-reversal related pair of Wannier functions will identically vanish.

Now consider the case of a Zeeman field $B_x$ but no N\'eel order ($N_y=0$). While $\mathcal T$ is clearly broken, inversion symmetry $\mathcal I$ is preserved. The spectrum, shown in Fig.~\ref{fig:zigzag}(c) by the blue curves, is no longer manifestly twofold degenerate, but also has energy gap. Note that the buckled chain with Zeeman field has a twofold rotation symmetry $\{\mathcal C_{2x} |\tfrac12\}$, such that the Hamiltonian $H_k$ block-diagonalizes (see Appendix~\ref{app:1D-chain}).The dashed and solid blue curves Fig.~\ref{fig:zigzag}(c) correspond to the energy bands in the two blocks, which are labeled by eigenvalues of $\{\mathcal C_{2x} |\tfrac12\}$. Inversion symmetry would seem to forbid a polarization $P_x$, but since polarization is only defined up to an integer multiple of $e$, the presence of $\mathcal I$ symmetry gives rise to two distinct classes 1D insulators with quantized polarization $P_x=0$ and $P_x=e/2$~\cite{Hughes:2011p245132,Miert2017}. Crucially, here we find that the 1D buckled chain with Zeeman field $B_x$ belongs to the second class of insulators, and thus has polarization $P_x=e/2$. A proof and detailed discussion is provided in Appendix~\ref{app:1D-chain}.

\begin{figure}
	\includegraphics[width=\columnwidth]{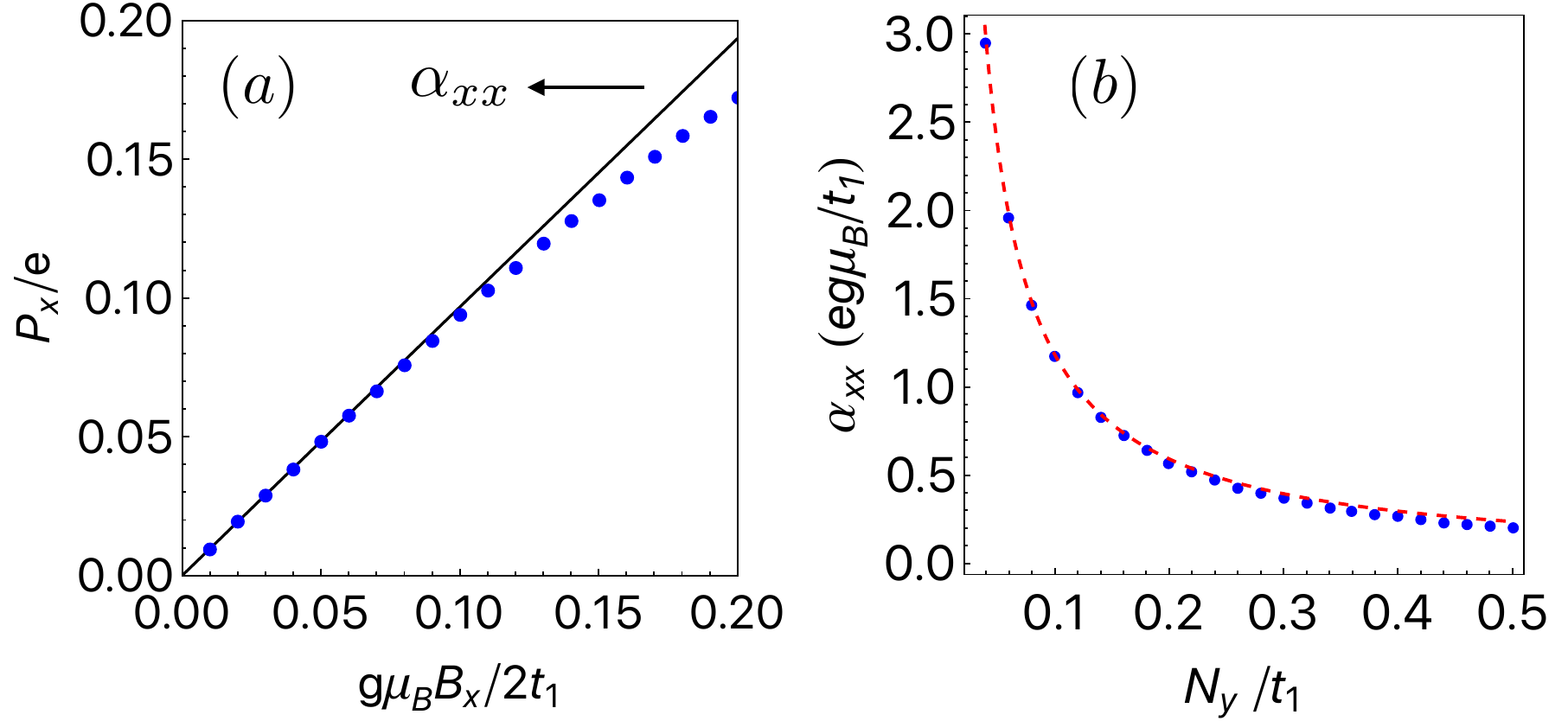}
	\caption{(a) Berry phase polarization $P_x$ of the 1D buckled chain model given by Eq.~\eqref{eq:H_k-1D-buckled} as a function of Zeeman field $B_x$. Blue dots represent calculated polarization at finite $B_x$ using the modern theory of polarization~\cite{Resta:1994p899}. Black line corresponds to $\alpha_{xx}$ calculated at $B_x=0$ using \eqref{eq:SMP-1D}. Here we have chosen $\lambda = 0.8t_1$ and $N_y=0.2t_1$. (b) Shown is $\alpha_{xx}$ as a function of $N^y$. Blue dots represent the result calculated using Eq.~\eqref{eq:SMP-1D} and the dashed red curve is approximate result from a continuum Dirac model at $k=\pi$. Here we have chosen $\lambda = 0.4t_1$.}
	\label{fig:zigzag_P}
\end{figure}

The footprint of the half-integer polarization of the insulating state in the presence of a Zeeman field is manifested in the SMP $\alpha_{xx}$, which is the quantity that precisely captures the polarization response to a Zeeman field. Applying Eq.~\eqref{eq:SMP-4band} to the 1D model defined by Eq.~\eqref{eq:H_k-1D-buckled} (with $B_x=0$), we find 
\be
\alpha_{xx} =  -\frac12 eg\mu_B \int \frac{dk}{2\pi} \frac{\lambda N_y c_k }{(N^2_y+4t^2_1c^2_{k/2}+\lambda^2s^2_k)^{3/2}}. \label{eq:SMP-1D}
\ee
In Fig.~\ref{fig:zigzag_P}(a) we show the calculated Berry phase polarization $P_x$ as a function of Zeeman field, obtained for $\lambda = 0.8t_1$ and $N_y=0.2t_1$ in Eq.~\eqref{eq:H_k-1D-buckled}. The solid black line represents the linear relation $P_x = \alpha_{xx} B_x$, with $\alpha_{xx} $ calculated from \eqref{eq:SMP-1D}, and demonstrates that $\alpha_{xx} $ indeed describes the linear magnetoelectric polarizability. 

The topological signature of the 1D model becomes apparent when evaluating $\alpha_{xx}$ as a function of N\'eel order parameter $N_y$, as is shown in Fig.~\ref{fig:zigzag_P}(b). As $N_y$ approaches zero, the polarizability $\alpha_{xx}$ diverges, which can be understood by expanding the integrand in \eqref{eq:SMP-1D} around $k=\pi$, where a massive Dirac model describes the dispersion. This yields the result
\be
\alpha_{xx}\approx  \frac{eg\mu_B}{2\hbar v}  \int \frac{dq}{2\pi} \frac{\lambda N_y }{(N^2_y+q^2)^{3/2}} = \frac{eg\mu_B\lambda}{4\pi\hbar v N_y} 
\ee
which, as evidenced by Fig.~\ref{fig:zigzag_P}(b), correctly captures the $\sim 1/N_y$ divergence of the full lattice calculation. 

\subsection{Two-legged Su-Schrieffer-Heeger ladder}
\label{ssec:SSH}

\begin{figure}[tbp]
	\includegraphics[width=1\columnwidth]{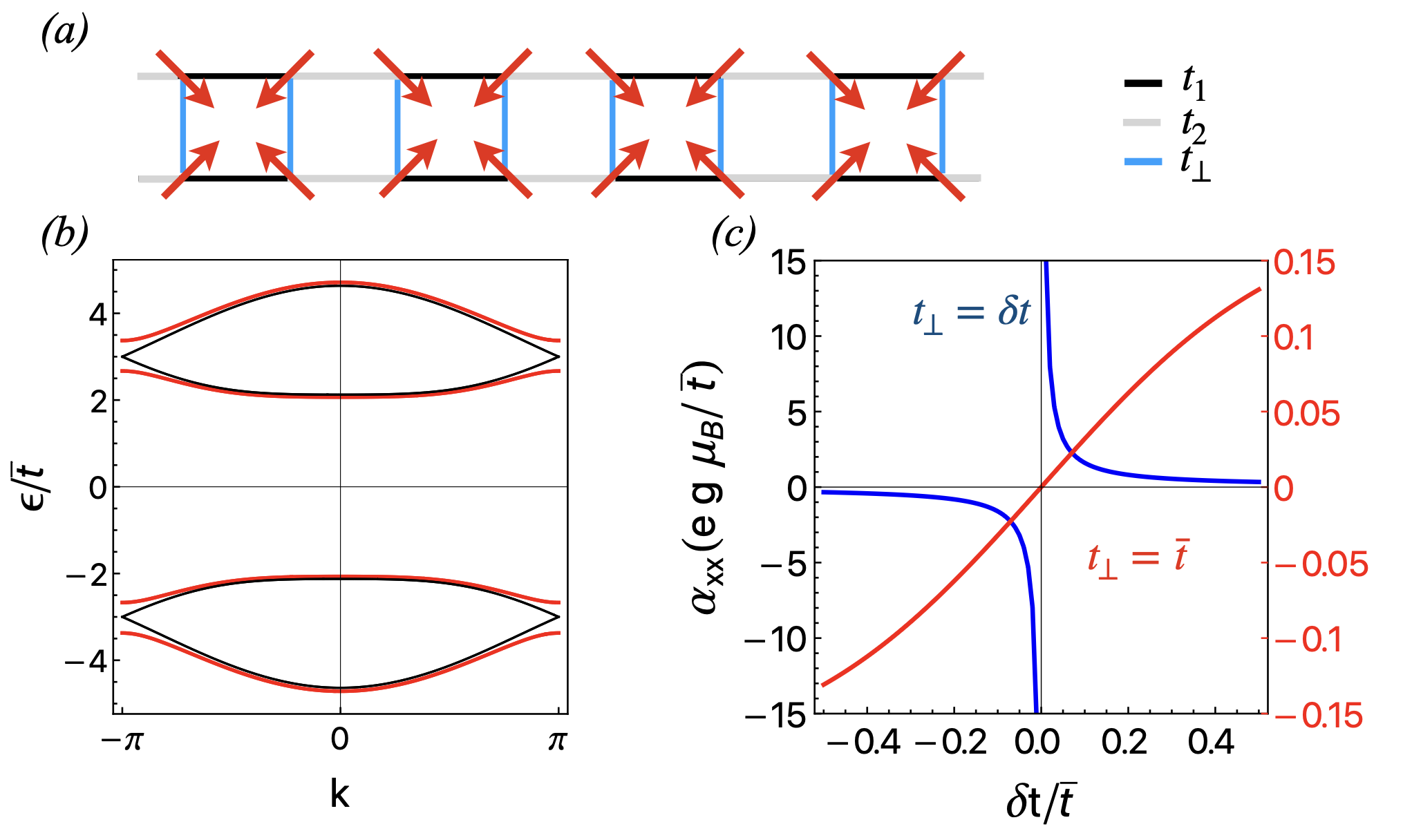}
	\caption{(a) Sketch of the two-legged Su-Schrieffer-Heeger ladder discussed in Sec.~\ref{ssec:SSH}. The center of the squares are centers of the product symmetry ${\mathcal I}{\mathcal T}$. (b) Energy spectrum for a vanishing interchain hopping amplitude $t_\perp$ and $\bar{t} = (t_1 + t_2)/2$ displaying a fourfold degeneracy (black thin lines) that is gapped out by Zeeman coupling (red thick lines). The Kondo coupling $J=3 \bar{t}$. (c) Diagonal magnetoelectric polarizability $\alpha_{xx}$ as a function of the bond density wave order parameter $\delta t = t_1-t_2$, assuming an interleg hopping amplitude $t_\perp=\delta t$ (blue line) and $t_\perp=\bar{t}$ (red line). In the former case the divergence as $\delta t \rightarrow 0$ signals the topological contribution to the magnetoelectric polarizability.}
	\label{fig:doubleSSH}
\end{figure}

The second model we consider corresponds to a 1D version of the double-Q spin model on the square lattice discussed in Sec.~\ref{sec:noncollinear}. The 1D variant of the double-Q spin model is shown in Fig.~\ref{fig:doubleSSH}(a). It can be viewed as a spinful two-legged Su-Schrieffer-Heeger (SSH) ladder model with a Kondo coupling between itinerant electron spins and an ordered configuration of local classical moments. As shown in Fig.~\ref{fig:doubleSSH}(a), the configuration of moments is chosen equal to the configuration of a unit layer of Fig.~\ref{fig:figvdbmodel}(a). In particular, within each leg the magnetic configuration has a ferromagnetic component (along ${\hat y}$) and antiferromagnetic component (along ${\hat x}$), and the ferromagnetic components are equal and opposite for the two legs. The alternating intraleg hopping amplitudes are denoted $t_{1}$ and $t_{2}$, and are crucial to obtain a finite magnetoelectric polarizability. Finally, we introduce an interleg hopping $t_{\perp}$ [see Fig.~\ref{fig:doubleSSH}(a)], which couples the two legs.

This SSH ladder model preserves the combined ${\mathcal I} \mathcal T$ symmetry when the inversion center is chosen as the center of the square. Therefore, all bands are twofold degenerate at all momenta. In addition, the model retains the two product symmetries ${\mathcal M}_{x} \mathcal T$ and ${\mathcal M}_{y} \mathcal T$, where the mirror planes intersect the center of the bonds. A symmetry-based Landau theory adapted from the theory developed in Sec.~\ref{sec:noncollinear} then shows that the only symmetry-allowed magnetoelectric coupling is of the form $\alpha_{xx} M_x P_x$. In what follows we examine the magnetoelectric coefficient $\alpha_{xx}$ microscopically and demonstrate that this model exhibits the same essential features as the zigzag chain model.

Consider first the situation in which the intraleg hopping amplitudes $t_{1,2}$ are assumed to be equal and the interleg hopping is set to zero, i.e., $t_{\perp}=0$. When the intraleg hoppings are equal ($t_1=t_2$) the system has an additional symmetry composed of a half translation and a spin rotation given by $ \mathcal U_{2y}\left\{ \mathbb{1} | \frac{1}{2}  \right\}$. Under this symmetry the electric polarization $P_x$ does not change, $P_x \rightarrow P_x$, while the magnetization $M_x$ is reversed, $M_{x} \rightarrow -M_{x}$. Hence, this additional symmetry forbids the appearance of any magnetoelectric coupling. The energy bands of SSH ladder model for $t_1=t_2$ and $t_{\perp}=0$ are shown in Fig.~\ref{fig:doubleSSH}(b) in black. Two manifolds of bands are centered at energies $-J$ and $J$, and the bands within each manifold cross at $k=\pi$, thus giving rise to a gapless phase at quarter and three-quarter filling. The gapless point at $k=\pi$ may be understood as follows. When $t_{\perp} = 0$ the legs of the ladder are uncoupled and may examined independently. Viewing each leg as an independent 1D chain, we observe that such a chain as an additional inversion symmetry with respect to the atomic sites. This inversion symmetry anticommutes with $ \mathcal U_{2y}\left\{ \mathbb{1} | \frac{1}{2}  \right\}$ at $k=\pi$, which implies a twofold degeneracy of all energy levels and explains band crossings shown in Fig.~\ref{fig:doubleSSH}(b) in the limit of uncoupled legs.

Let us now consider the effect of making the two intraleg hoppings different, i.e., $t_1 \neq t_2$, as well as the effect of introducing a Zeeman field $B_x$. As in Sec.~\ref{sec:noncollinear}, we define $\delta t = t_1-t_2$ and $\bar{t} = (t_1 + t_2)/2$ to parametrize the alternating hoppings. Both $\delta t$ and the Zeeman field $B_x$ lead to an energy gap at quarter (and three-quarter) filling; the case of a nonzero Zeeman field is shown in Fig.~\ref{fig:doubleSSH}(b) in red. The two resulting insulating states are different, however. The insulating state resulting from a nonzero Zeeman field $B_x$ (with $\delta t =0$) has a quantized electronic contribution to the polarization, as explicitly shown in Appendix~\ref{app:sshladderpolarization}. Specifically, the total electronic contribution to the polarization of the two-legged ladder is given by $ e/2~\textrm{mod}~e$. Instead, the insulating state produced by nonzero $\delta t$ does not have a half-integer quantized polarization. This is fully analogous to the situation encountered in the case of the spin-orbit coupled zigzag chain model, where the N\'eel ordered insulator was trivial (vanishing polarization) and the ``Zeeman insulator'' had a nontrivial half-integer quantized polarization.

As for the buckled zigzag chain, the half-integer polarization of the Zeeman field insulator leaves a topological footprint in the SMP. To demonstrate this, we show the magnetoelectric polarizability $\alpha_{xx}$ as a function of $\delta t$ (for $B_x=0$) in Fig.~\ref{fig:doubleSSH}(c). The blue curve corresponds to $t_{\perp} = \delta t$ and exhibits a divergence as $\delta t \rightarrow 0 $, which is the analog of the divergence shown in Fig.~\ref{fig:zigzag_P}(b) as $N_z \rightarrow 0$. It reflects the fact that, as $\delta t$ decreases, the system responds more strongly to a Zeeman field due to half-integer quantized polarization at $\delta t =0$. Importantly, we also find that this topological signature is absent when the interleg hopping amplitude $t_{\perp}$ is set to $\bar t$, as is shown by the red curve in Fig.~\ref{fig:doubleSSH}(c). In this case the SMP shows behavior consistent with simple symmetry arguments and vanishes when $\delta t =0$. The reason is that $t_{\perp}$ competes with a Zeeman field and produces a trivial insulating state without half-quantized polarization.

We thus find that the SSH ladder model provides a spin-orbit free realization of the essential features found in the buckled zigzag chain model. The spin magnetoelectric polarizability can exhibit a divergence as the symmetry which forbids a magnetoelectric effect are restored. This divergence is a signature of nontrivial crystalline topology in 1D and the associated half-quantized electronic polarization.

\section{Discussion and Conclusion \label{sec:conclusions}}

In this final section we summarize and discuss the main results presented in this paper. The discussion will emphasize three aspects in particular. The first concerns the way in which the topological magnetoelectric effect revealed in this work relates to the known response theory of topological phases in 1D, 2D, and 3D. The second concerns connections to the notion of anomalous spin and also to topological insulator surface states. The third concerns possible material candidates.

\subsection{Pattern of electromagnetic response}
\label{ssec:patterns-EM}

The central result of this work is the topological contribution to the spin magnetoelectric polarizability of 2D antiferromagnets. In general, antiferromagnets which admit a linear magnetoelectric response break inversion ($\mathcal I$) and time-reversal symmetry ($\mathcal T$), but preserve their product ($\mathcal I \mathcal T$), and can be classified in terms of higher order magnetoelectric multipole moments (i.e. magnetoelectric monopole, toroidal dipole, magnetic quadrupole). The topological contribution to the spin magnetoelectric polarizability occurs in the special class of Dirac antiferromagnets, by which we mean insulating antiferromagnets described by a gapped Dirac theory~\footnote{Dirac antiferromagnets should therefore be distinguished from antiferromagnetic Dirac semimetals, which are antiferromagnetic semimetallic gapless states protected by crystal symmetries.}. As the analysis of the two microscopic models shows, this class includes spin-orbit coupled antiferromagnets with collinear N\'eel order, but also spin-orbit-free antiferromagnets with noncollinear order. The key characteristic shared between these two representative models is that the essential electronic structure of the magnetic insulator is adequately captured by a continuum Dirac model.  

The origin of the topological magnetoelectric effect is directly linked to this Dirac structure and can be traced back to the electromagnetic response theory of Dirac semimetals in 2D~\cite{Ramamurthy:2015p085105}. To demonstrate this, we have detailed how the topological spin magnetoelectric polarizability---which describes the Berry-phase polarization response to a Zeeman field---can be understood from examining the effect of a Zeeman field on the Dirac crossings. The Zeeman field enters the Dirac description as a pseudo-gauge field and shifts the Dirac nodes in momentum space, thus creating a family of effectively 1D Hamiltonians with half-integer quantized polarization. As explained in Sec.~\ref{ssec:em-dirac}, this observation establishes a connection between the magnetoelectric response of Dirac antiferromagnets in 2D and the quantized electromagnetic response of 1D crystalline insulators. The connection between $\mathbb{Z}_2$ classifications of crystalline insulators in 1D and magnetoelectric response was further explored in Sec.~\ref{sec:1D} in the context of microscopic toy models of 1D antiferromagnets. 

It is worth considering in more detail how the magnetoelectric response in two dimensions fits into a pattern of electromagnetic responses of crystalline solids in one, two, and three dimensions. Here we discuss this pattern from the perspective effective actions, which provides a unifying framework for understanding the responses. 

The polarization of a 1D insulator is described by an effective action of the form~\cite{Qi:2008p195424,Hughes:2011p245132,Turner:2012p165120,Ramamurthy:2015p085105}
\be
S_{\text{1D}}[A_\mu] = \frac{e P_1}{2}\int dxdt ~ \epsilon^{\mu\nu} F_{\mu\nu}, \label{eq:S_1D}
\ee
where $ F_{\mu\nu} = \partial_\mu A_\nu-\partial_\nu A_\mu$ is electromagnetic field strength tensor ($\mu,\nu=0,1$ in 1D), and $P_1$ is the polarization. When a coordinate-inverting crystalline symmetry is present---inversion symmetry is an obvious example---the polarization must satisfy $P_1 = -P_1~\text{mod}~\mathbb{Z}$, which implies that $P_1 - n = 0,1/2$, where $n$ is an integer. The two possible values $0$ and $1/2$ define the $\mathbb{Z}_2$ classification of 1D crystalline insulators, and Eq.~\eqref{eq:S_1D} captures the electromagnetic response of such insulators (i.e., the quantized polarization)~\cite{Hughes:2011p245132}.
 
A generalization of \eqref{eq:S_1D} occurs in three dimensions and describes the electromagnetic response of 3D topological insulators. The effective action reads as
\be
S_{\text{3D}}[A_\mu] = \frac{e^2 P_3}{16 \pi h}\int d^3xdt ~ \epsilon^{\mu\nu\lambda \rho} F_{\mu\nu}F_{\lambda\rho},\label{eq:S_3D}
\ee
where now $P_3$ is referred to as the magnetoelectric polarization. Time-reversal symmetry imposes a quantization condition on $P_3$ such that $P_3 - n = 0,1/2$, where $n$ is again an integer, and the two possible values $0$ and $1/2$ define the $\mathbb{Z}_2$ classification of 3D time-reversal invariant insulators. (Note that other symmetries can have similar implications~\cite{Turner:2010p241102,Hughes:2011p245132,Turner:2012p165120,Zhang:2019p206401}.) 
The effective action \eqref{eq:S_3D} describes the topological magnetoelectric effect of 3D topological insulators mentioned in the introduction. This can be seen more directly by rewriting $\epsilon^{\mu\nu\lambda \rho} F_{\mu\nu}F_{\lambda\rho}/8 = \bE\cdot \bB $. 

The two effective actions in Eqs.~\eqref{eq:S_1D} and \eqref{eq:S_3D} thus describe the topological electromagnetic response of insulators in 1D and 3D with a $\mathbb{Z}_2$ classification protected by discrete (crystalline) symmetries. In 3D, the response corresponds to the topological magnetoelectric effect mentioned in the introduction, which has been a subject of great interest since it was first pointed out. The polarization and magnetoelectric responses are topological in the sense that they are quantized in units of natural constants, i.e., the electric charge and $e^2/h$, respectively.

The topological magnetoelectric effect revealed in this work does not derive from a response theory of the form \eqref{eq:S_1D} or \eqref{eq:S_3D}, i.e., the response theory of insulators, but instead derives from the response of topological semimetals. More specifically from the response theory of Dirac semimetals in 2D~\cite{Ramamurthy:2015p085105} (see Sec.~\ref{ssec:em-dirac}). The effective action governing the response takes the form
\be
S_{\text{2D}}[A_\mu] = \frac{e }{4 \pi }\int d^2xdt ~ \epsilon^{\mu\nu\lambda } b_\mu F_{\nu\lambda }, \label{eq:S_2D}
\ee
from which the response equation \eqref{eq:P_i-a_j} may be obtained by variation with respect to the electric field. The effective action features the 3-vector $b_\mu = (b_0,\bb)$, where, as explained in Sec.~\ref{ssec:em-dirac}, $\bb$ corresponds to the separation of the Dirac nodes in momentum space (we comment on the time component $b_0$ below).  As mentioned in Sec.~\ref{ssec:em-dirac},  the response described by \eqref{eq:S_2D} can be understood as the response arising from a stack or collection of 1D systems with a response described by \eqref{eq:S_1D}. This can now be seen by comparing Eq.~\eqref{eq:S_2D} with Eq.~\eqref{eq:S_1D}, and noting that the former can be viewed as the result of promoting the 1D topological response to 2D, with the help of $b_\mu$. Since $b_\mu$ clearly has dimensions, the 2D response is more appropriately referred to as a quasitopological response, which is 
a typical feature for the electromagnetic response of topological semimetals~\cite{Ramamurthy:2015p085105}. 

This brief review of electromagnetic responses thus shows how the response theory underlying the topological magnetoelectric effect in 2D fits into a pattern of responses of topological semimetals in various dimensions, which themselves derive from the topological response of insulators in lower dimensions. Note, however, that it is far from obvious that the effective action \eqref{eq:S_2D} describes a magnetoelectric effect. Indeed, in the context in which it was applied and discussed by Ramamurthy and Hughes~\cite{Ramamurthy:2015p085105} it did not. In the present context, its manifestation as a magnetoelectric response follows from the identification of $\bb$ with the magnetic Zeeman field $\bB$, as is described by Eq.~\eqref{eq:a_x-a_y}, and the way it enters the theory of an antiferromagnetic 2D Dirac semimetal.
Therefore, since $\bb$ is linearly related to $\bB$, the effective action \eqref{eq:S_2D} describes a magnetoelectric coupling $\sim EB$. Since the magnetoelectric coupling relies on the Zeeman coupling of magnetic field and spin, it is a spin magnetoelectric coupling. This is different from the topological magnetoelectric coupling in 3D, given by Eq.~\eqref{eq:S_3D}, which relies on the effect of the magnetic field on the orbital motion of Bloch electrons. 

\subsection{Connection to anomalous spin and TI surface states \label{ssec:spin}}

In this work we have viewed and discussed the spin magnetoelectric effect from the perspective of a polarization response to an applied Zeeman field. The Maxwell relation given by Eq.~\eqref{eq:maxwell} implies a complementary perspective: a spin magnetization response to an electric field. Therefore, when viewed from this perspective, our analysis implies that Dirac antiferromagnets also exhibit a topological spin magnetization response to an electric field. 

In the broader context of spin magnetization responses to electric fields, it is interesting to relate the magnetoelectric tensor to the anomalous spin polarizability $\Upsilon_{ij} $~\cite{Garate:2009p134403,Kurebayashi:2014p211,Dong:2020p066601,Xiao:2023p166302}. The latter describes a spin density response to an electric field, typically of interest in metals, and is referred to as ``anomalous'' in analogy with the anomalous velocity~\cite{Xiao:2023p166302,Xiao2010}, also typically relevant to metals. It can be directly inferred from Eqs.~\eqref{eq:SMP} and \eqref{eq:M_i-s_i} that the magnetoelectric polarizability and the anomalous spin polarizability are simply related as
\be
 \alpha_{ij}=\frac{g\mu_B}{\hbar}\Upsilon_{ij} .
\ee
Furthermore, it follows from Eq.~\eqref{eq:SMP-4band} that for the class of four-band models discussed in Sec.~\ref{ssec:tight-binding} the $\bk$-space anomalous spin polarizability density associated with the occupied bands is given by 
\be
\Upsilon_{ij} (\bk)=  e\hbar  \frac{\partial_i\bn    \times \ehat_j \cdot \bn }{2 \varepsilon_\bk^3}. \label{eq:anomalous-spin}
\ee
This simplified expression for the anomalous spin polarizability density may be compared to similar expressions for the $\bk$-space Berry curvature $\Omega_{ij} (\bk)$ applicable to generic two-bands models~\cite{Graf:2021p085114}. Clearly, the Berry curvature associated with the occupied bands of of Hamiltonian system in Eq.~\eqref{eq:H_k} vanishes, but Eqs.~\eqref{eq:anomalous-spin} reveals that its cousin, the anomalous spin polarizability density, is generally non-vanishing and captures a different band-geometric property of the occupied states. The analysis of Sec.~\ref{ssec:em-dirac} exposes the way in which the integrated spin polarizability density, in the case Dirac antiferromagnets, reflects a topological property.

A connection between (anomalous) spin density responses and topology was first pointed out by Garate and Franz in the context of topological insulator surface states coupled to ferromagnetic order~\cite{Garate:2010p146802}. Our analysis enables us to revisit their result within a model for a topological insulator thin film with a top and bottom surface, both coupled to a ferromagnetic layer. When the magnetization of the adjacent ferromagnetic layers is oriented along the direction of the surface normal for both surfaces (i.e., in an antiferromagnetic and inversion-breaking fashion), the Hall conductivities of the two surfaces cancel, but the magnetoelectric and anomalous spin responses \emph{add}. Instead, when the direction of magnetization of, for instance, the bottom surface is reversed, giving rise to an overall ferromagnetic state, the Hall conductivities of the two surfaces add. This leads to a $\nu =1$ state. The contributions to the magnetoelectric and anomalous spin response from each surface cancel in this case.

\subsection{Connections to materials}

Next, we turn to possible connections with experimentally relevant materials. When considering material candidates, the most important requirement is that the essential electronic properties are described by a Dirac theory, since that is the origin of the topological magnetoelectric effect. The buckled square lattice model considered in Sec.~\ref{sec:square} suggests that (layered) materials with a nonsymmorphic space group are promising candidates, owing to the symmetry protection of Dirac crossings provided by the nonsymmorphic symmetries. Hence, this specific model with space group $P4/nmm$ is a representative of a larger class of crystalline systems with nonsymmorphic space groups and symmetry-enforced Dirac points at high symmetry points on the BZ boundary. The space group analysis presented in Appendix~\ref{app:P4/nmm} provides a recipe for determining which systems and materials exhibit such symmetry-enforced Dirac points in the non-magnetic state. Development of antiferromagnetic order can then give rise to magnetoelectric effect discussed in this work.

An example of a magnetic material with the same (nonmagnetic) space group as the buckled square lattice model is CuMnAs~\cite{Wadley:2013p2322}, which has attracted considerable attention as a platform for antiferromagnetic spintronics~\cite{Wadley:2016p587,Smejkal:2017p106402,Godinho:2018p4686}, as well as phenomena arising from inversion-breaking antiferromagnetism~\cite{Wang:2021p277201}. CuMnAs can be thought of as a layered structure, in which the magnetic Mn within each layer form a buckled square lattice. In this sense it is well described by the model discussed in Sec.~\ref{sec:square} \cite{Smejkal:2017p106402}. 

To make further contact with experimentally relevant settings, let us consider order-of-magnitude estimate of the predicted magnetoelectric effect. Since the (quasitopological) spin magnetoelectric polarizability of a Dirac model only depends on the Fermi velocity (and not on the gap), an estimate for the quasitopological polarizability may be obtained by considering a typical Fermi velocity of the order of $v \simeq 10^5 m/s$. This yields an estimated polarizability of the order of $2.2~10^{-14}$~A$^2$~s/N, assuming a bare $g=2$ Land\'e $g$-factor.  We note that estimating the Fermi velocity as $v \simeq 10^5 m/s$ is equivalent to estimating the length scale $a=\hbar/(m_e v) $ as $a\simeq 1$ \r{A}. It is interesting to compare this estimated value of the (quasitopological) spin magnetoelectric polarizability with the values of the total magnetoelectric polarizability obtained by very recent density functional theory calculations for a large class of 2D magnets~\cite{Mangeri:2026p015011}. For the purpose of comparison it is important to note that the magnetoelectric polarizability as defined in Ref.~\onlinecite{Mangeri:2026p015011} differs from the magnetoelectric polarizability considered in our manuscript by a factor $\mu_0$, since the former has been defined as $\alpha_{ij}=\partial P_i / \partial H_j$. With this definition $\alpha_{i j}$ has units of seconds, and Ref.~\onlinecite{Mangeri:2026p015011} finds that the majority of the screened 2D magnets display a global magnetoelectric polarizability of the order of $10^{-3}$ attoseconds. Our estimate of the quasitopological spin magnetoelectric polarizability corresponds to $2.5~10^{-2}$ attoseconds, which is one order of magnitude larger.  

In the context of experimentally accessible systems it is worth mentioning that the zigzag chain model is realized in a number of three-dimensional but quasi-1D coupled-chain compounds. Examples include NdRu$_2$Al$_10$ and TbRu$_2$Al$_10$~\cite{Yanase:2014p014703}, and also Ce$_3$TiBi$_5$~\cite{Hayami:2022p123701}. Such systems provide interesting venues for examining the magnetoelectric properties of the 1D models discussed in Sec.~\ref{sec:1D}.

\subsection{Concluding remarks}

We conclude with a number of final brief remarks. First, we emphasize that the magnetoelectric coupling discussed in this work does not rely on the presence of spin-orbit coupling, but can also occur in magnets with no or weak spin-orbit interaction. Our analysis of the spin-orbit-free double-$Q$ antiferromagnet on the square lattice demonstrates this. Second, this manuscript focuses specifically and exclusively on the spin magnetoelectric response, i.e., the polarization response to a Zeeman field. One might wonder about orbital contribution to the magnetoelectric response, which refers to the polarization induced by the effect of the magnetic field on the orbital motion of the electrons. Such a magnetoelectric response was studied in detail in Ref.~\onlinecite{Essin:2010p205104}. The microscopic expression for the orbital magnetoelectric polarizability obtained in Ref.~\onlinecite{Essin:2010p205104} applies to three-dimensional systems, however. The models of reduced dimension studied in this work do not give rise to an orbital contribution to magnetoelectric polarizability.

Finally, the effective action of Eq.~\eqref{eq:S_2D} describes a response more general than what has been considered here. Indeed, the magnetoelectric effect this work has focused on is associated with the spatial components of $b_\mu = (b_0,\bb)$, namely $\bb$. It is then natural to ask what the response associated with time-component $b_0$ is. Furthermore, variation of Eq.~\eqref{eq:S_2D} with respect to the vector potential suggests that there are charge and current responses associated with spatial and temporal variations of $\bb$. Since $\bb$ is linearly related to the Zeeman field, this in turn suggest that charge and current responses in Dirac antiferromagents can be manipulated and controlled by spatial variations of the Zeeman field. This may be achieved by proximity coupling to magnetic textures via heterostructure engineering. We anticipate that these are fruitful directions to be explored in future works.

\section*{Acknowledgements}

We have benefited greatly from insightful discussions with Daniel Agterberg, Cristian Batista, Charles Kane, Rafael Fernandes, Denys Makarov, Jeroen van den Brink, and David Vanderbilt. We thank Daniel Agterberg in particular for useful comments concerning material realizations. We also gratefully acknowledge Harini Radhakrishnan for collaboration on a related project. 
J.W.F.V. was supported by the U.S. Department of Energy (DOE) under Early Career Award No. DE-SC0025632.
P.G. acknowledges financial support from the Italian Ministry of University and Research (MUR) under
the National Recovery and Resilience Plan (NRRP), Call PRIN 2022, funded by the European Union
NextGenerationEU, PNRR-M4C2, I1.1 project 2022HTPC2B (TOTEM)- CUP B53D23004210006 and
PNRR-M4C2, I1.3 Extended Partnership NQSTI - PE0000023 CUP B53C22004180005. C.O. acknowledges partial support by the Italian Ministry of Foreign Affairs and International Cooperation PGR12351 (ULTRAQMAT) and from PNRR MUR Project No. PE0000023-NQSTI (TOPQIN).

\appendix

\section{On the units of magnetoelectric polarizability \label{app:units}}

This Appendix provides a brief analysis of the units of the spin magnetoelectric polarizability (SMP). The goal is to demonstrate explicitly that the SMP given by Eq.~\eqref{eq:SMP} has the same units as the orbital magnetoelectric polarizability introduced in Refs.~\onlinecite{Essin:2010p205104} and \onlinecite{Malashevich:2010p053032}.

Consider first the integrand of Eq.~\eqref{eq:SMP}. It is useful to define an energy scale of the order of the bandwidth as $W \equiv (\hbar/a)^2/2m_e$, where $a$ is the lattice constant and $m_e$ is the electron mass. The dimension of the numerator comes from the derivative of the Hamiltonian (with respect to wave vector $k_i$) and the dimension of the denominator is determined by the band energy. Since both the energy and the Hamiltonian can be expressed in units of $W$, and since the integration measure has units of $1/a^d$, the integral has units of
\be
a^{-d} \times  \frac{a W}{W^2} =  \frac{a^{1-d}}{W}.
\ee
It follows that the SMP $\alpha_{ij}$ has units of
\be
 \frac{e\mu_B a^{1-d}}{W},
\ee
which may be simplified by observing that
\be
\frac{\mu_B}{W}   =\frac{2m_e }{ (\hbar/a)^2}  \frac{e\hbar}{2m_e} =a^2 \frac{e}{\hbar} . 
\ee
The SMP can thus be expressed in units of
\be
 \frac{e\mu_B a^{1-d}}{W} = a^{3-d} \frac{e^2}{\hbar} , 
\ee
which is indeed equal to $e^2/\hbar$ in 3D, as found for the orbital magnetoelectric polarizability~\cite{Essin:2010p205104,Malashevich:2010p053032}. That the topological Chern-Simons piece of the orbital magnetoelectric polarizability should have units of $e^2/\hbar$ was reported earlier in Refs.~\onlinecite{Essin:2009p146805,Qi:2008p195424}. We observe that in 2D the SMP has units of $a e^2/\hbar$, thus implying a dependence on the material-specific lattice constant.

\section{Generalized four-band model \label{app:four-band}}

This Appendix describes the derivation of a formula for the SMP applicable to the case of general four-band models. Sec.~\ref{ssec:tight-binding} considers a particular class of four-band models, namely those which can be expanded in the particular set of Pauli matrices featured in Eq.~\eqref{eq:H_k}. Here we consider a slightly more general class, where the explicit form of the generating matrices is left unspecified. The formula for the SMP derived here is therefore representation-independent. 

We consider four-band Hamiltonians of the general form
\be
H = h^\mu \Gamma_\mu, \label{app:H-Gamma}
\ee
where $\Gamma_\mu$ are a set of five $4\times 4$ Hermitian matrices which anticommute, i.e., 
\be
\{\Gamma_\mu,\Gamma_\nu \} = 2\delta_{\mu\nu}, \label{app:Gamma-commute}
\ee
and $h^\mu$ is a five-component real vector, which may also be written simply as $\bh$. Note that these five components are generally functions of momentum $\bk$, but we suppress (the obvious) momentum dependence in this Appendix for simplicity. Four-band Hamiltonians which preserve the product of inversion and time-reversal can always be written in this form. In Eq.~\eqref{eq:H_k}, the specific representation of $\Gamma$-matrices is given by $\Gamma_\mu=(\tau^x,\tau^y,\tau^z\bsigma)$. 

The energy spectrum of \eqref{app:H-Gamma} is easily obtained and consists of two branches $\varepsilon_{\pm }$ given by
\be
\varepsilon_{\pm } = \pm \varepsilon, \qquad \varepsilon = \sqrt{\bh^2}= |\bh|.
\ee
The Hamiltonian may be written in terms of the projectors $P_\pm$ onto the two corresponding eigenspaces as
\be
H =  \varepsilon ( P_+- P_- ) ,
\ee
with $P_\pm$ given by
\be
P_\pm =  \frac12(\mathbb{1} \pm H/\varepsilon )=   \frac12(\mathbb{1} \pm h^\mu \Gamma_\mu/\varepsilon ). \label{app:projectors}
\ee
Note that none of this requires specifying the form of the $\Gamma$-matrices; all these properties follow from the anticommutation relation. 

To compute $\alpha_{ij}$ we first note that Eq.~\eqref{eq:integrand-projector} remains valid in this general case, and the task is therefore to determine $ \text{Im}  \text{Tr}[ P_- \hat v_i P_+ \hat s_j ] $ with projectors given by \eqref{app:projectors}. To evaluate the trace we need the appropriate expressions for $ \hat v_i $ and $ \hat s_j$, and for the velocity one clearly has
\be
\hat v_i =\hbar^{-1} \partial_i H = \hbar^{-1}(\partial_i h^\mu )\Gamma_\mu.
\ee
The spin components $\hat s_j $ can very generally be expanded as
\be
\hat s_j  = s^\mu_j \Gamma_\mu + \omega^{\mu\nu}_j \Omega_{\mu\nu},
\ee
where $ \Omega_{\mu\nu}$ are defined as
\be
 \Omega_{\mu\nu} = \frac{1}{2i} [\Gamma_\mu,\Gamma_\nu].
\ee
Since $ \Omega_{\mu\nu}=- \Omega_{\nu\mu}$ there are ten such matrices, and together with the five $\Gamma$-matrices these span the set of $4\times 4$ Hermitian matrices. It must therefore be possible to write the spin operators in this way. In anticipation of the result below we may note that the components $s^\mu_j $ are irrelevant, since terms involving $s^\mu_j $ do not contribute to the trace. 

Inserting the form of $ \hat v_i $ and $ \hat s_j$, as well as the projectors, we then use trace formulas for products of $\Gamma$-matrices to arrive at the result
\be
 \text{Im}  \text{Tr}[ P_- \hat v_i P_+ \hat s_j ] = -4 \frac{h^\mu (\partial_i h^\nu) \omega^{\mu\nu}_j}{\hbar \varepsilon},
\ee
and substituting this into the expression for $\alpha_{ij} $ yields
\be
\alpha_{ij} =  eg\mu_B \int \frac{d^2\bk}{(2\pi)^2} \frac{2 h^\mu (\partial_i h^\nu) \omega^{\mu\nu}_j}{\hbar \varepsilon^3}.
\ee
This is the final result for the more general class of four-band Hamiltonians. 

We may check that this expression reduces to \eqref{eq:SMP-4band} when the $\Gamma$-matrices are chosen as $(\tau^x,\tau^y,\tau^z\bsigma)$ and $\hat s_j = \hbar \sigma^j/2$. Take $j=x$, for instance. In this case we have 
\be
\omega^{45}_x = -\omega^{54}_x = \frac{\hbar}{4},
\ee
with all other components zero, since
\be
\omega^{45}_x \Omega_{45} +\omega^{54}_x\Omega_{54} =\frac{\hbar}{4i} [\tau^z\sigma^y,\tau^z\sigma^z] = \frac{\hbar}{2} \sigma^x. 
\ee
We then have 
\begin{align}
\frac{2 h^\mu \partial_i h^\nu \omega^{\mu\nu}_x}{\hbar \varepsilon^3}  &= \frac{ h^4 \partial_i h^5 - h^5 \partial_i h^4}{2 \varepsilon^3}\\
& = \frac{ n_y \partial_i n_z - n_z \partial_i n_y}{2 \varepsilon^3} \\
&= \frac{\bn   \times\partial_i\bn \cdot \ehat_x  }{2 \varepsilon^3},
\end{align}
which is precisely Eq.~\eqref{eq:SMP-4band}.

\section{Space group $P4/nmm$ \label{app:P4/nmm}}

This Appendix provides a detailed analysis of the space group symmetries of the buckled square lattice model featured in Sec.~\ref{sec:square}. The goal is to establish the space group symmetry-enforced fourfold band degeneracies at the high symmetry points on the Brillouin zone boundary. The essence of such analysis was reported in Ref.~\onlinecite{Young:2015p126803}; here we revisit this analysis to emphasize the algebraic relations between space group elements. 


To demonstrate that all energies at the boundary high-symmetry points must be fourfold degenerate, we seek pairs of space group symmetries which have the following properties: (i) the representations of these symmetries in the space of Bloch bands anticommute at a given high symmetry point, and (ii) the representation of at least one of these symmetries squares to $+1$ at that point (i.e., square to the identity). These properties may be directly inferred from the space group algebra. Together with time-reversal symmetry, the presence of two symmetries which satisfy these criteria implies the fourfold degeneracy of all energy levels. 

In what follows $\mathcal R$ will denote a rotation $2\pi$, which has Bloch state representation $-1$ ($+1$) in the case of spinful (spinless) electrons. Since here we include the effect of SOC the spinful case applies. We further make use of the fact the representation of translations $\{ \mathbb{1} | \ba \}$ in the space of Bloch states is $e^{-i \bk \cdot \ba}$.  The high symmetry points of interest are $M=(\pi,\pi)$, $X_1=(\pi,0)$, and $X_2=(0,\pi)$.

We now examine pairs of symmetries. (In a slight abuse of language we will say that a symmetry squares to $\pm1$ at a particular high symmetry point, even though this is technically a statement about the representation of the symmetry in the space of Bloch states.)

\noindent 
{\it The glide mirror $\{\mathcal M_{z} | \tfrac12 \tfrac12 \}$ and inversion $\{\mathcal I | 00 \} $.} Squaring each of these symmetries yields
\be
\{\mathcal M_{z} | \tfrac12 \tfrac12 \}^2 = \mathcal R \{ \mathbb{1} | 11 \}, \quad  \{\mathcal I | 00 \}  ^2 =\{ \mathbb{1} | 00 \}  ,  \label{eq:2D-algebra-1a}
\ee
from which it follows that $\{ M_{z} | \tfrac12 \tfrac12 \}$ squares to $+1$ at $X_1$ and $X_2$, but squares to $-1$ at $M$. The commutation relation can be established by noting that
\be
\{ \mathcal M_{z} | \tfrac12 \tfrac12 \}\{ I | 00 \}=  \{ \mathbb{1} | 11 \} \{ \mathcal I | 00 \}\{ \mathcal M_{z} | \tfrac12 \tfrac12 \}, \label{eq:2D-algebra-1b}
\ee
from which it follows that $\{ \mathcal M_{z} | \tfrac12 \tfrac12 \}$ and $\{\mathcal  I | 00 \} $ anticommute at $X_1$ and $X_2$. It then follows that all energies must be fourfold degenerate at $X_1$ and $X_2$.

{\it The twofold screw $\{ \mathcal C_{2x} | \tfrac12 0  \}$ and the twofold screw $\{ \mathcal C_{2y} |0 \tfrac12   \}$.} Squaring each of these symmetries yields
\be
\{ \mathcal C_{2x} | \tfrac12 0  \}^2 = \mathcal R \{ \mathbb{1} | 10 \}, \quad  \{ \mathcal C_{2y} | 0\tfrac12  \}^2 =  \mathcal R \{ \mathbb{1} | 01 \}  ,  \label{eq:2D-algebra-2a}
\ee
from which it follows that $\{ \mathcal C_{2x} | \tfrac12 0  \}$ squares to $+1$ at $X_1$ and $M$, whereas $\{ \mathcal C_{2y} |0 \tfrac12   \}$ squares to $+1$ at $X_2$ and $M$. Evaluating the commutation relation we find
\be
\{ \mathcal C_{2x} | \tfrac12 0  \}  \{ \mathcal C_{2y} | 0\tfrac12  \}  = \mathcal R  \{ \mathbb{1} | 1\bar 1 \} \{ \mathcal C_{2y}  | 0\tfrac12  \}\{ \mathcal C_{2x} | \tfrac12 0  \} . \label{eq:2D-algebra-2b}
\ee
from which it follows that $\{ \mathcal C_{2x} | \tfrac12 0  \}$ and $\{ \mathcal C_{2y} |0 \tfrac12   \}$ anticommute at $M$, but commute at $X_1$ and $X_2$.

{\it The twofold screw $\{\mathcal C_{2x} | \tfrac12 0  \}$ and the mirror $\{ \mathcal M_{y} | 0\tfrac12  \} $.} Squaring each of these symmetries yields
\be
\{\mathcal C_{2x} | \tfrac12 0  \}^2 = \mathcal R \{ \mathbb{1} | 10 \}, \quad  \{\mathcal M_{y} | 0\tfrac12  \}^2 =  \mathcal R \{ \mathbb{1} | 00 \}  ,  \label{eq:2D-algebra-3a}
\ee
from which it follows that $\{\mathcal C_{2x} | \tfrac12 0  \}$ squares to $+1$ at $X_1$ and $M$, whereas $\{\mathcal M_{y} | 0\tfrac12  \} $ always squares to $-1$. Evaluating the commutation relation we find
\be
\{\mathcal C_{2x} | \tfrac12 0  \}  \{\mathcal M_{y} | 0\tfrac12  \} =\mathcal R  \{ \mathbb{1} | 0\bar 1 \} \{\mathcal M_{y} | 0\tfrac12  \}\{\mathcal C_{2x} | \tfrac12 0  \} . \label{eq:2D-algebra-3b}
\ee
from which it follows that $\{\mathcal C_{2x} | \tfrac12 0  \}$ and $\{\mathcal M_{y} | 0\tfrac12  \} $ anticommute at $X_1$, but commute at $X_2$ and $M$.

{\it The glide mirror $\{\mathcal M_{z} | \tfrac12 \tfrac12 \}$ and the mirror $\{\mathcal M_{1\bar 1} | 00 \} $.} Note that by $\mathcal M_{1\bar 1}$ here we mean a mirror reflection in a plane with normal direction $[1\bar 1]$. Squaring each of these symmetries yields
\be
\{ \mathcal M_{z} | \tfrac12 \tfrac12 \}^2 = \mathcal R \{ \mathbb{1} | 11 \}, \quad  \{\mathcal M_{1\bar 1} | 00 \} ^2 =  \mathcal R \{ \mathbb{1} | 00 \}  ,  \label{eq:2D-algebra-4a}
\ee
from which it follows that $\{\mathcal M_{z} | \tfrac12 \tfrac12 \}$ squares to $+1$ at $X_1$ and $X_2$, whereas $\{\mathcal M_{1\bar 1} | 00 \} $ only leaves $M$ invariant and squares to $-1$ at $M$. Evaluating the commutation relation we find
\be
\{ \mathcal M_{z} | \tfrac12 \tfrac12 \}  \{ \mathcal M_{1\bar 1} | 00 \}= \mathcal R  \{ \mathcal M_{1\bar 1} | 00 \}\{ \mathcal M_{z} | \tfrac12 \tfrac12 \}   \label{eq:2D-algebra-4b},
\ee 
from which it follows that $\{ \mathcal M_{z} | \tfrac12 \tfrac12 \}$ and $\{ \mathcal M_{1\bar 1} | 00 \} $ anticommute at $M$.

{\it The  twofold screw $\{ \mathcal C_{2x} | \tfrac12 0  \}$ and inversion $\{ \mathcal I | 00 \} $.} Squaring each of these symmetries yields
\be
\{ \mathcal C_{2x} | \tfrac12 0  \}^2 = \mathcal R \{ \mathbb{1} | 10 \}, \quad  \{ \mathcal I | 00  \}^2 =  \{ \mathbb{1} | 00 \}  ,  \label{eq:2D-algebra-3a}
\ee
from which it follows that $\{ \mathcal C_{2x} | \tfrac12 0  \}$ squares to $+1$ at $X_1$ and $M$. Evaluating the commutation relation we find
\be
 \{ \mathcal I | 00  \}\{ \mathcal C_{2x} | \tfrac12 0  \}   =   \{ \mathbb{1} | \bar 10 \} \{ \mathcal C_{2x} | \tfrac12 0  \}  \{ \mathcal I | 00  \} \label{eq:2D-algebra-3b}
\ee
from which it follows that $\{ \mathcal C_{2x} | \tfrac12 0  \}$ and $\{ \mathcal I | 00 \} $ anticommute at $X_1$ and  $M$.

\section{Landau theory of magnetoelectric coupling \label{app:landau}}

To determine the magnetoelectric response of a N\'eel ordered state within a Landau theory framework, we consider a Landau free energy of the general form $\mathcal F = \mathcal F_P +\mathcal F_M + \mathcal F_N + \mathcal F_{P,M,N} $. The first two terms are given by
\be
\mathcal F_P + \mathcal F_M = \frac{1}{2\chi_e}\bP^2 + \frac{1}{2\chi_m}\bM^2 - \bE \cdot \bP- \bB \cdot \bM,
\ee
where $\bP$ is the polarization and  $\bM$ is the magnetization. Here $\chi_e$ and $\chi_m$ are the electric and magnetic susceptibilities, and both $\bP$ and $\bM$ are coupled to their respective conjugate fields, i.e., the electric field $\bE$ and magnetic field $\bB$. As is well-known, a theory of just $\mathcal F_P + \mathcal F_M $ is sufficient to describe the polarization response to an electric field and the magnetization response to a magnetic field. 

The contribution $\mathcal F_N $ is an expansion in the N\'eel order parameter, which describes the development of magnetic order. Its explicit form depends on the type of N\'eel order, easy-plane or easy-axis, which cannot be determined from symmetry arguments alone. If $\mathcal F_N $ is meant to describe easy-axis ordering (as is considered in the main text), it takes the form 
\be
\mathcal F_N = a N^2_z +  bN^4_z ,
\ee
whereas easy-plane order would be described by a free energy given by
\be
\mathcal F_N = a(N^2_x+N^2_y) + b_1(N^4_x+N^4_y) + b_2 N^2_xN^2_y.
\ee
In keeping with the analysis of the main text, here we assume N\'eel order along the $z$ direction. In this case, the final term $\mathcal F_{P,M,N} $, which describes the trilinear couplings, can be expressed as
\be
\mathcal F_{P,M,N} = \bar{\alpha}_{ij} P_i M_j N_z.
\ee

Minimization of the free energy $\mathcal F$ with respect to $\bP$ and $\bM$ yields the two coupled equations
\begin{align}
P_i &= \chi_e E_i  - \chi_e\bar{\alpha}_{ij}M_j N_z, \\
M_i &= \chi_m B_i  -\chi_m \bar{\alpha}_{ji}P_j N_z.
\end{align}
Solving for $P_i $ and $M_i$ while only retaining the lowest order linear terms in $N_z$ finally yields
\begin{align}
P_i &\simeq \chi_e E_i  - \alpha_{ij}B_j , \\
M_i &\simeq \chi_m B_i  -\alpha_{ji}E_j ,
\end{align}
where we have introduced $ \alpha_{ij}= \chi_e\chi_m \bar{\alpha}_{ij}N_z$.

\section{Dirac point moving in the double-$Q$ spin model
\label{app:dirac-shift-double-Q}}

In this appendix we show that the magnetic field-induced shift of the Dirac cone in the double-$Q$ spin model of Sec.~\ref{sec:noncollinear}  is not a specific feature of the low-energy model but  is a symmetry-enforced process. 

Let us specialize to the spin configuration depicted in Fig.~\ref{fig:figvdbmodel}(a) and consider the effect of a Zeeman coupling due to a magnetic field along the $\hat{x}$ direction.  A finite magnetic field  along the ${\hat x}$ axis breaks the fourfold rotation symmetry as well as the combined ${\mathcal I} \mathcal T$ symmetry of the model, but preserves the glide reflection symmetry  $\left\{ M_x | 0 \frac{1}{2} \right\}$. In the two-dimensional BZ there are two ``glide" lines, $k_x=0$ and $k_x=\pi$,  that are mapped onto themselves under the glide reflection symmetry.
The electronic bands are then labeled by the glide reflection eigenvalues. Importantly, while at the $\Gamma$ and $X$ point the glide eigenstates are non-degenerate, at the $Y$ and $M$ points there is an effective Kramers degeneracy.  This is because the system preserves another antiunitary symmetry given by the composition of the mirror $M_y$ (with the mirror plane at the center of the bonds) and time-reversal. This antiunitary symmetry squares to one and does not guarantee degeneracies. However, the Bloch Hamiltonian also commutes with $\left\{ M_x | 0 \frac{1}{2} \right\} M_y \mathcal{T}$ that squares to minus one at $Y$ and $M$ points and hence predicts Kramers degeneracies. We therefore find that, similarly to the situation encountered in transition metal dichalcogenides in the 1T$^{\prime}$ structure~\cite{Muechler2016}, the twofold degeneracies at even fillings are symmetry protected: the splitting of the fourfold degenerate Dirac point in a pair of twofold degeneracies is the only symmetry-allowed process in the presence of an external magnetic field.
This guarantees the robustness of the topological term in the SMP as well as in the spin multipolization.

\section{Buckled 1D chain \label{app:1D-chain}}

In this Appendix we collect details of the analysis of the 1D zigzag chain model. We begin by demonstrating that the space group symmetries of the nonmagnetic lattice mandate fourfold degenerate energy levels at $k=\pi$. Following the recipe detailed in Appendix~\ref{app:P4/nmm}, we examine the algebraic relations between space group elements to identify a pair of symmetries sufficient to guarantee fourfold degeneracies.  

To this end, consider first the square of the rotations, which are given by
\be
 \{ \mathcal C_{2x} | \tfrac12 \}^2 = \mathcal R \{ \mathbb{1} | 1 \}, \;\;  \{ \mathcal C_{2z} | 0 \} ^2 =  \{ \mathcal C_{2y} | \tfrac12  \} ^2 = \mathcal R . \label{eq:rotation-algebra}
\ee
Next, consider the mirror reflections, for which we find
\be
 \{ \mathcal M_{x} | \tfrac12 \}^2 =   \{ \mathcal M_{z} | 0 \} ^2 = \mathcal R ,\;\;  \{ \mathcal M_{y} | \tfrac12  \} ^2 = \mathcal R \{ \mathbb{1} | 1 \} . \label{eq:mirror-algebra}
\ee
From these relations we conclude the following that at $k=\pi$ the (representations of the) symmetries $ \{ \mathcal C_{2x} | \tfrac12 \}$ and $ \{ \mathcal M_y | \tfrac12 \}$ must square to $+1$. (Recall that, as in Appendix~\ref{app:P4/nmm}, $\mathcal R=-1$ for spinful fermions.)

Next, consider the commutation relation between $ \{ \mathcal C_{2x} | \tfrac12 \}$ and $ \{ \mathcal M_y | \tfrac12 \}$. We find
\be
\{ \mathcal C_{2x} | \tfrac12 \}  \{ \mathcal M_{y} | \tfrac12 \} =\mathcal R   \{ \mathcal M_{y} | \tfrac12 \} \{ \mathcal C_{2x} | \tfrac12 \} , \label{eq:commutation}
\ee
which implies that $ \{ \mathcal C_{2x} | \tfrac12 \}$ and $ \{ \mathcal M_y | \tfrac12 \}$ anticommute at $k=\pi$. We conclude that we have identified two point group symmetries which square to $+1$ and mutually anticommute at $k=\pi$. As a result, all bands must be fourfold degenerate at $k=\pi$, which implies in particular that the fourfold linear crossing of the zigzag chain model at $k=\pi$ is symmetry-enforced.

We now turn to the allowed magnetoelectric response in the N\'eel ordered state. To determine the response for given orientation of the N\'eel vector, we follow the Landau theory approach of Sec.~\ref{ssec:square-landau} and construct the trilinear couplings between polarization, magnetization, and N\'eel order. For the 1D zigzag chain we find
\begin{multline}
\mathcal F_{P,M,N} = (\beta_{xxy} P_xM_x  + \beta_{yyy} P_y  M_y + \beta_{zzy} P_z M_z) N_y \\
+ (\beta_{xyx}  P_xM_y+ \beta_{yxx} P_y M_x)N_x \\
+ (\beta_{yzz}  P_yM_z+ \beta_{zyz} P_z M_y)N_z. \label{app:1D-trilinear}
\end{multline}
This is in agreement with previous work which discussed aspects of the allowed response of zigzag chains~\cite{Yanase:2014p014703,Hayami:2015p064717,Hayami:2022p123701,Yatsushiro:2022p155157,Hayami:2023p094106}. Here we are primarily interested in bulk polarization and therefore focus on terms involving $P_x$. (A polarization $P_y$ might be defined as charge imbalance between the two sublattices.) The specific case of a polarization response to a Zeeman field discussed in the main text assumed N\'eel order along $y$, which is described by the term $\beta_{xxy} P_xM_xN_y$ in \eqref{app:1D-trilinear}. A similar result may be obtained by choosing the N\'eel vector along $x$ and the Zeeman field along $y$.

We proceed with a more detailed analysis of the case discussed in the main text (N\'eel vector along $\hat y$; Zeeman field along $\hat x$). The Hamiltonian of Eq.~\eqref{eq:H_k-1D-buckled} can be rewritten as
\be
H_{k} = f_x  \tau^x  +n_z \tau^z\sigma^z +  n_y \tau^z \sigma^y +b_x \sigma^x, \label{app:H_k-1D}
\ee
which takes the general form of Eq.~\eqref{eq:H_k}, but with an additional Zeeman term. Here we have set $f_x = -2 t_1 c_{k/2}$ and 
\be
(n_z,n_y)= (- \lambda s_k, N_y),\quad b_x = -\frac12 g\mu_BB_x.
\ee
To examine the symmetries in the presence of N\'eel order and the Zeeman field, consider first the case of N\'eel order ($n_y\neq 0$) but no Zeeman field ($b_x=0$). N\'eel order breaks $ \{ \mathcal I |  \tfrac12 \}$ and $\mathcal T$, but preserves the product symmetry $\{ \mathcal I |  \tfrac12 \} \mathcal T$. N\'eel order furthermore preserves the twofold rotation symmetries $\{ \mathcal C_{2z} | 0\}$ and $\{ \mathcal C_{2x} | \tfrac12 \}$. These rotation symmetries imply that the Hamiltonian (with $b_x=0$) satisfies the relations
\be
\tau^x\sigma^z H_k  = H_{-k}\tau^x\sigma^z , \quad \tau^x\sigma^x H_k = H_{k}\tau^x\sigma^x ,
\ee
respectively. 

Instead, when N\'eel order is absent ($n_y= 0$) but a Zeeman field is applied ($b_x\neq 0$) inversion symmetry $\{ \mathcal I |  \tfrac12 \}$ is preserved (while $\mathcal T$ is manifestly broken). Whereas the twofold rotation symmetry $\{ \mathcal C_{2z} | 0\}$ is broken in the presence of the Zeeman field, the twofold screw rotation $\{ \mathcal C_{2x} | \tfrac12 \}$ is preserved. 

It thus follows that when N\'eel order and the Zeeman field are both present, the twofold rotation $\{ \mathcal C_{2x} | \tfrac12 \}$ remains a symmetry and $H_k$ of Eq.~\eqref{app:H_k-1D} commutes with $\tau^x\sigma^x$. As a result, writing the Hamiltonian in a basis of eigenstates of $\tau^x\sigma^x$ will render $H_k$ block diagonal. We find that the two blocks are given by 
\be
H^\pm_k =   \begin{pmatrix} n_z &  \pm f_x +b_x-in_y \\ \pm f_x +b_x+in_y & -n_z  \end{pmatrix} ,
\ee
where $\pm$ corresponds to the eigenvalues $\pm 1$ of $\tau^x\sigma^x$. The spectrum in each sector is given by the two branches $+\varepsilon^\pm$ and $-\varepsilon^\pm$, where
\be
\varepsilon^\pm = \sqrt{n^2_y+n^2_z+(b_x \pm f_x)^2}. \label{app:eps_pm}
\ee

In the main text we have shown the effect of N\'eel order and a Zeeman field on the Dirac crossing at $k=\pi$ in Fig.~\ref{fig:zigzag}(c). The effects are easily extracted from \eqref{app:eps_pm}.  Expanding in small momentum $q$ around $k=\pi$ we find that the spectrum for nonzero $N_y $ takes the form
\be
\varepsilon^\pm_q = \sqrt{N^2_y+( t^2_1+ \lambda^2)q^2},
\ee
which corresponds to a massive Dirac fermion with velocity $\hbar v =\sqrt{t^2_1+ \lambda^2}$ and mass $N_y$. Note that the spectrum is identical in the two rotation eigenvalues sectors (labeled by $\pm$). 

In the absence of N\'eel order but with nonzero Zeeman field the dispersion close to $k=\pi$ is given by
\be
\varepsilon^\pm_q = \sqrt{ \lambda^2q^2 +(b_x \pm t_1 q)^2}.
\ee
This also describes the dispersion of a massive Dirac fermion, but shifted in momentum and in opposite directions for the two eigenvalue sectors. The dispersion is shown in Fig.~\ref{fig:zigzag}(c) by the blue curves, with dashed and solid curves corresponding to the two rotation eigenvalue sectors. 

The main text mentions the fact that the antiferromagnetic ``N\'eel insulator'' (i.e., $b_x=0$ but nonzero $N_y$) has zero Berry phase polarization $P_x$, whereas the ``Zeeman insulator'' (i.e., $N_y=0$ but nonzero $b_x$) represents a $\mathbb{Z}_2$ topological phase in 1D protected by inversion symmetry, with quantized polarization $P_x=e/2$ mod $e$. To demonstrate this, we construct a family of 1D insulators, parametrized by an angle $\theta$, which interpolate between these two insulators. Specifically, we construct a family of Hamiltonians $H^\eta_k (\theta) = \bd^\eta \cdot \brho$, where $\brho = (\rho^x,\rho^y,\rho^z)$ are Pauli matrices and $\eta=\pm$, with $\bd^\eta = (d^\eta_x,d^\eta_y,d^\eta_z)$ given by
\begin{align}
d^\eta_x & = -\eta 2t_1 \cos(k/2) +b_x\sin\theta ,\\
d^\eta_y & = N_y \cos\theta, \\
d^\eta_z & = - \lambda \sin k.
\end{align}
When $\theta = 0$ the Hamiltonian $H^\eta_k (\theta)$ describes the N\'eel insulator, and when $\theta = \pi/2$ the Hamiltonian describes the Zeeman insulator. For all intermediate values of $\theta$ this family of Hamiltonians maintains a gapped spectrum. The difference in polarization $\Delta P_x$ between the two insulators of interest is given by \cite{Resta:1992p51,King-Smith:1992p1651,Vanderbilt:1993p4442,Resta:1994p899}
\be
\Delta P_x = e \int^{\pi/2}_{0}d\theta \int^\pi_{-\pi}\frac{dk}{2\pi} ( \Omega^+_{k\theta}+\Omega^-_{k\theta}),
\ee
where the curvatures $\Omega^\eta_{k\theta}$ can be expressed as
\be
\Omega^\eta_{k\theta}= \frac{\bd^\eta \cdot \partial_k\bd^\eta\times \partial_\theta \bd^\eta}{2 |\bd^\eta |^3}.
\ee
Here $\partial_k = \partial/ \partial k$ and $\partial_\theta = \partial/ \partial \theta$. Inserting the expressions for $\bd^\eta$, we find
\be
\bd^\eta \cdot \partial_k\bd^\eta\times \partial_\theta \bd^\eta = \frac12 N_y \lambda (b_x c_k +\eta 2s_\theta c^3_{k/2}),
\ee
and
\be
|\bd^\eta | = \sqrt{N^2_yc^2_\theta+\lambda^2 s^2_k+(b_xs_\theta \mp t_1c_{k/2})^2},
\ee
where, as in the main text, we have abbreviated $c_k \equiv \cos k$, $c_\theta \equiv \cos \theta$, and $s_\theta \equiv \sin \theta$. Assuming that $t_1>0$ we then find the result
\be
\Delta P_x = \text{sgn}(\lambda N_yb_x)\frac{e}{2}.
\ee

The fact that the Zeeman insulator is an inversion symmetry protected $\mathbb{Z}_2$ topological phase is also reflected in the fact that, as $b_x$ is increased, a gap closing transition necessarily occurs at $k=0$. This transition occurs when $b_x \pm f_x=0$, and therefore only occurs in one of the two rotation eigenvalues sectors. At the transition two bands with opposite inversion eigenvalues cross, thus changing the parity of the product of inversion eigenvalues of the occupied bands.

\section{Quantized polarization in the two-legged SSH ladder \label{app:sshladderpolarization}}

In this Appendix we demonstrate that in the presence of inversion symmetry, {\it i.e.} for vanishing interleg hopping amplitude $t_{\perp}$, the two legs composing the SSH ladder lead to a 
half-integer electronic contribution to the polarization $e/2~\textrm{mod}~e$.
Let us start by recalling the elementary band representations~\cite{Zak1982} for exponentially localized and symmetric Wannier functions in inversion symmetric chains. In particular, since we are interested in the magnetoelectric coupling at one-quarter and three-quarter filling, we only need the elementary band representation for Wannier functions with maximal Wyckoff positions whose stabilizer group contains inversion. We have that for the Wannier function with center of charge with coordinate $x=0$ in the inversion-symmetric unit cell, the two inversion eigenvalues at the center and edge of the 1D BZ will both be $+1$. On the contrary, the Wannier function with center of charge with coordinate $x=a/2$ will be characterized by inversion eigenvalue $+1$ at the $\Gamma$ point and invergion eigenvalue $-1$ at the $X$ point of the BZ. Let us now fix the origin of both the ladders leg at the lattice site where the classical spin has positive $\hat{x}$ component [see the inset of Fig.~\ref{fig:doubleSSH}(a)]. Remember that the inversion centers for the two uncoupled legs and $t_1=t_2$ correspond to the spin sites. We can now compute the inversion eigenvalues by simply noticing that the inversion operator for each leg can be written as ${\mathcal I}=\textrm{diag}\left(1, \mathrm{e}^{ik}\right) \otimes \sigma_0$. It is straightforward to see that at one-quarter filling the two Wannier functions of the two legs will be centered on our chosen unit cell origin for a magnetic field $B_x>0$ and instead will be centered at the maximal Wyckoff position with $x=a/2$ for $B_x<0$. This result could be also anticipated by considering the atomic limit $t_{1,2} \rightarrow 0$ where the Hamiltonian reduces to the local exchange term $J~{\bf S} \cdot \boldsymbol{\sigma}$ augmented with the Zeeman coupling $B_x \sigma_x$. 

Having established the center of charge of the two Wannier functions for the two independent legs, we can now compute the electronic contribution to the total polarization of the two-legged SSH ladder. In doing so, we recall that for finite values of $t_\perp$,   the absence of any electronic contribution of polarization at zero magnetic field is protected by ${\mathcal I} \mathcal{T}$, which requires fixing the origin of the unit cell at the centers of the squares. 
As we are interested in difference of polarization, we need to evaluate the electric polarization for the inversion symmetric case at finite fields with the same origin choice. In this case, the center of charges have coordinate $x=\mp a/4~\textrm{sign}(B_x)$ and hence the contribution to the polarization will be 
$P_x=\mp e/2~\textrm{sign}(B_x)$. 
Thus, the electronic contribution to the polarization always assumes half-integer values.

\end{document}